# Church's thesis is questioned by new calculation paradigm

Hannes Hutzelmeyer

**Summary:**

Church's thesis claims that all effecticely calculable functions are recursive. A shortcoming of the various definitions of recursive functions lies in the fact that it is not a matter of a syntactical check to find out if an entity gives rise to a function. **Eight new ideas** for a precise setup of arithmetical logic and its metalanguage give the proper environment for the construction of a special **computer**, the **ARBACUS** computer. Computers do not come to a necessary halt; it is requested that **calculators** are constructed on the basis of computers in a way that they always come to a halt, then all calculations are effective. The **ARBATOR** is defined as a calculator with **two-layer-computation**. It allows for the calculation of all primitive recursive functions, but **multi-level-arbation** also allows for the calculation of other **arbative** functions that are not primitive recursive. The **new paradigm** of calculation does not have the above mentioned shortcoming. The defenders of Church's thesis are challenged to show that **exotic** arbative functions are recursive and to put forward a recursive function that is not arbative. A construction with **three-tier-multi-level-arbation** that includes a diagonalisation leads to the extravagant yet calculable **Snark-function** that is not arbative. As long as it is not shown that all exotic arbative functions and particularily the Snark-function are **arithmetically representable** Gödel's first incompleteness sentence is in limbo.

## 1. Introduction

### 1.1 On the way to recursive functions

It is not easy to read textbooks on mathematical logics - this is at least my impression. It is strange that a topic that should illuminate our thinking defies a simple access. When I started to work on it some years ago I decided for myself to do better and bring more order and add some beauty. It was both a matter of content **and** aesthetics. I did not start rightout with the idea that there might be some open questions in mathematical logics, although I always had a strange feeling towards some Gödel-type so-called theorems [2] [7], but who has not such a feeling if self-reference is involved ? And there was always this strange animal called Church's thesis [4] [5]. One would not allow for a thesis in mathematics, why allow for one in metamathematics or **supramathematics**\*\*) (supra: meta-meta). The use of axioms within a theory is something else, as it is not claimed that they **are** intrinsically true. Every theorem of an axiomatic theory really reads: if the axioms are true then this and that is true. If one finds a system where the axioms are actually true one knows that the theorems are true. If not, it is just a *Glasperlenspiel*. There are conjectures e.g. like the Goldbach-conjecture, but nobody would call a sentence that depends either on the truth or the falsity of the Goldbach-conjecture a theorem. Only in the if-then-form it could be a theorem.

The time between 1926 and 1936 must have been very exciting until there finally was a sustainable concept of effectively calculable functions that included Ackermann-functions [1] : around 1934 the Princeton circle of Church, Kleene and Gödel (and Herbrand) introduced **minimisation** as an effective procedure complementing **primitive recursions**, which on their part consist of **straight-recursion**\*) and composition, starting from identity, projections and succession functions. By the way it was technically very simple to introduce the concept of minimisation with respect to arithmetical respresentations that are in the center of Gödel's work. It was much simpler than straight-recursion, where one needed Gödel's ingenious beta-function-technique. After that everything looked fine, except for the **ontology problem**\*) (as I call it) that gave some people some headache, but obviously not too much in the last seventy years; more of this in the next section. In the years after 1936 various competing methods for effectively calculable functions have been put forward, the following list is not complete: Turing, Markov, lamda-calculus, Abacus, Register and so on. But all of them turned out to be equivalent definitions of recursive functions. And they all have a catch of the sort: you cannot tell in general whether a machine comes to a halt or if a function has at least one value zero or so.

### 1.2 Church's thesis and two theses of Gödel

As all the attempts to construct effectively calculable functions have turned out to lead to the same result this was considered as good evidence for **Church's thesis**: all effectively calculable functions are recursive. To my knowledge there has been no successfull attack on Church's calculability thesis, including hypercomputer concepts [8b] . And this is very important as some famous supratheorems depend on the truth of Church's calculability thesis. I call a proven sentence **of** a mathematical system a theorem and a proven metasentence\*) **about** a mathematical system a metatheorem and a proven suprasentence\*\*) **about** metamathematical systems a supratheorem\*). This brings up the immediate question: how can you call something a theorem or a metatheorem or a supratheorem if it depends on a thesis. Properly it has to be called a thesis too (that is why I have used in the summary and in section 1.1 the word "so-called"). And it does not help if somewhere in a first chapter it is written "under the asumption of Church's calculability thesis" or if one keeps repeating the mantra "assuming Church's calculability thesis", if one calls the outcome a theorem or metatheorem or supratheorem. The laymen readers take it as what you have **called** it. E.g. Gödel's so-called first incompleteness theorem (it is not a theorem but a suprasentence in the first place) really is **Gödel's first incompleteness thesis** as it depends on Church's calculability thesis.

---

\*) a star is attached if a word is given a new or special meaning, e.g. **thesis** is used for supralanguage sentences and **conjecture** for language and metalanguage sentences.   \*\*) two stars are attached to all words that I have coined newly.

And by the way: Gödel-type suprasentences say something about mathematics and nothing but: there are no other infinite language systems outside mathematics that one can reasonably talk about. Insofar they do not lend themselves for general philosophy. But whether authors explain it properly or not they start ranting about the consequences for science, philosophy, life in general and alleged limitations of the human mind. This is not my field.

Assume for the moment that Church's calculability thesis would turn out to be false, as it can happen with theses, otherwise they would not be theses. Some parts of the general public would perhaps maliciously point at mathematicians and claim that this is just another field where professors like to quarrel. It is just of intellectual comfort to respond that everything was only said under well-stated conditions.

What is the reason that such a problem can arise in mathematics, shouldn't it be free from eventual flaws. The answer is: one has to be precise, the problem does not arise **in** mathematics but only when one talks **about** mathematics. When one talks in mathematics (or even metamathematics) one usually has a well-defined system like group theory or real functions and no deep ontological problems, e.g. there are individuals, sets, mappings and predicates, sentences and formulae et cetera; and the mathematicians prove the truth of certain sentences. A sentence that starts with "for all" usually has a pretty good meaning. Perhaps this picture of mathematics is a little too romantic, but with a grain of salt that is what mathematics is all about.

When one talks about mathematics and metamathematics, that is when one talks in supramathematics**,** the situation is completely different. The fantasy and the creativity of mathematician seems to be without limits and they keep inventing all sorts of systems. A suprasentence that starts with "for all" is something completely different from normal mathematics, as it may comprise those systems that have not yet been invented, but that may be invented by future mathematicians, the domain is open. So it is quite natural that things like Church's calculability thesis exist and you should enjoy them, because they may provide an interesting area to work on.

In the following I will present some results on my work on Church's calculability thesis. I have yet to explain what I mean by ontology problems in connection with it. In a system of recursive functions you know what you mean by numbers, formulae or sentences, but you have not such a clear notion what a recursive function is. You cannot state "for all recursive functions" without problems as - roughly speaking - recursive functions are defined as programs that halt. As there is no general criterion for the halting of programs you have no criterion if a given program is a recursive function. There are certain classes of recursive functions, e.g. the one that is called primitive, with which you can do beautiful mathematics, but there always remains the Damokles-sword of nonhalting .

If one is not interested in Church's calculability thesis per se but rather on the important application in the proofs of so-called Gödel-type metatheorems one can overcome the ontological discussion as these only need Church's calculability thesis insofar as it is used in the metatheorem that all recursive functions are **arithmetically representable** [7] by logical formulae that use nothing but zero, succession, addition and multiplication (usually "arithmetically" is left away). In the following I will therefore use a weaker thesis that I (posthumously) call **Gödel's calculability thesis**: all effectively calculable functions are arithmetically representable.

This is much more convenient: suppose somebody has shown that Church's calculability thesis is false as he has produced a non-recursive effectively calculable function. If this function happens to be arithmetically representable no problem with Gödel-type theses occur. So you better check for Gödel's calculability thesis first. If you find an effectively calculable function that is not arithmetically representable Gödel's first incompleteness thesis is false. If it is unknown whether it is arithmetically representable or not, Gödel's first incompleteness thesis is in limbo.

## 2. Language environment

*In this chapter I sketch the environment that is necessary for the introduction of what I call the concrete calcule* <u>NU</u> *of decimal arbation\*\*⁾ natural numbers. This concrete calcule will turn out to be a real gold-mine, which will be exploited subsequently.*

### 2.1 Abstract and concrete calcules

My desire for a clearer view on mathematical logics could not be done without some new methods, both for content and notation. I did not hesitate to redefine some expressions and even to invent some new expressions. The textbook authors have invented new terminology too. I found their special characters pretty ugly and sometimes poorly documented; Cutland [6] may forgive me, that I quote notation of p. 241-245 as an example with strange scripts, arrows et cetera. As I am not affiliated with any organisation I felt completely free to follow new paths. This is the advantage of the independent private scholar. As this publication serves the purpose to distribute some **new results** I cannot go into all the details and I will only sketch some concepts as it is usually done in scientific magazine contributions, where the learned reader will understand them nevertheless. Some examples will help. A textbook is in work and will be published in some time. I start off with the first of eight ideas:

> (idea 1)    **abstract and concrete** calcules.

The name **calcule**\*\*⁾ was chosen as I mean something that may be called **calculus** in Latin or **Kalkül** in German; in English, however, **calculus** is already used for the theory of functions of real numbers. A calcule is a language system, it consists of sentences that are formed according to some syntax rules.

An **abstract calcule**\*⁾ is a formal system, it does not talk about anything. It starts with a list of sentences that are called **axioms**. The axioms and those sentences that can be obtained via logical **deduction** are called valid. The sentences of an abstract calcule can be valid, invalid or indefinite. The deeper logical meaning of an abstract calcule is that it allows to state the truth of some if-then-sentences where the if-clause states the existence (in whatever sense) of certain entities that fulfill some rules.

A **concrete calcule**\*⁾ talks about a **codex**\*⁾. A codex consists of **individuals** (finite **strings**\*⁾ of characters of a finite alphabet and a decidable equality relation) that are formed according to some syntax rules. Furthermore a codex can include the precise description of **calculation** procedures for some functions and relations through a **calculator**\*⁾. I call a machine a calculator if it halts for all **programs** with all possible **inputs**, whereas a **computer**\*⁾ is a machine that may or may not halt computing when given a certain program with a certain input. So far this is just wording, when it gets to the real description of calculators one must have a guarantee that no non-halting situations can occur. With this definition every calculation\*⁾ is effective and if something is calculable\*⁾ it is effectively calculable ("effectively calculable" then becomes a pleonasm). A computer computes, a calculator calculates a result ( in German: "ein Rechner rechnet, ein Kalkulator <u>be</u>rechnet ein Ergebnis"). Due to the reference to a codex a concrete calcule is **not a formal system**.

Sentences of a **concrete calcule** are true or false. The general method of finding the truth of basic sentences of concrete calcules is yet to be investigated. I give it the name **demonstration** as opposed to deduction. Once you have true sentences in a concrete calcule you can start using deduction for more true sentences.

Therefore a **proof** is either a deduction or a demonstration. In this publication I restrict my view of mathematics to abstract and concrete calcules and do not go into the question whether there are other meaningful fields of mathematical work.

## 2.2 Hierarchy of languages

In the context of this publication there are basically two very different types of languages

- proposition languages, which consist of sentences that can be true in some sense or not
- command languages of computer programs.

I start with proposition languages and state

> (idea 2)   **Mencish-Funcish** hierarchy of precise languages up to supra-tier.

If a language talks about another language it is a metalanguage relative to this language. For abstract and concrete calcules I introduce the language **Funcish**\*\*) (short for Functum-language). I do not use common language as its metalanguage, but rather another precisely defined language that I call **Mencish**\*\*) (short for Meta-Funcish). Metacalcules are precisely defined but they are not formal systems. A calcule is given a name that refers to its individual sort: in this publication the abstract calcule alpha and the concrete calule NU. They are formulated in Funcish and talked about in Mencish. Mencish is talked about in common language, which is English (or at least, what I, being of German tongue, consider to be English). Section 4.2 will define the command language **A0**\*\*) for the codex.

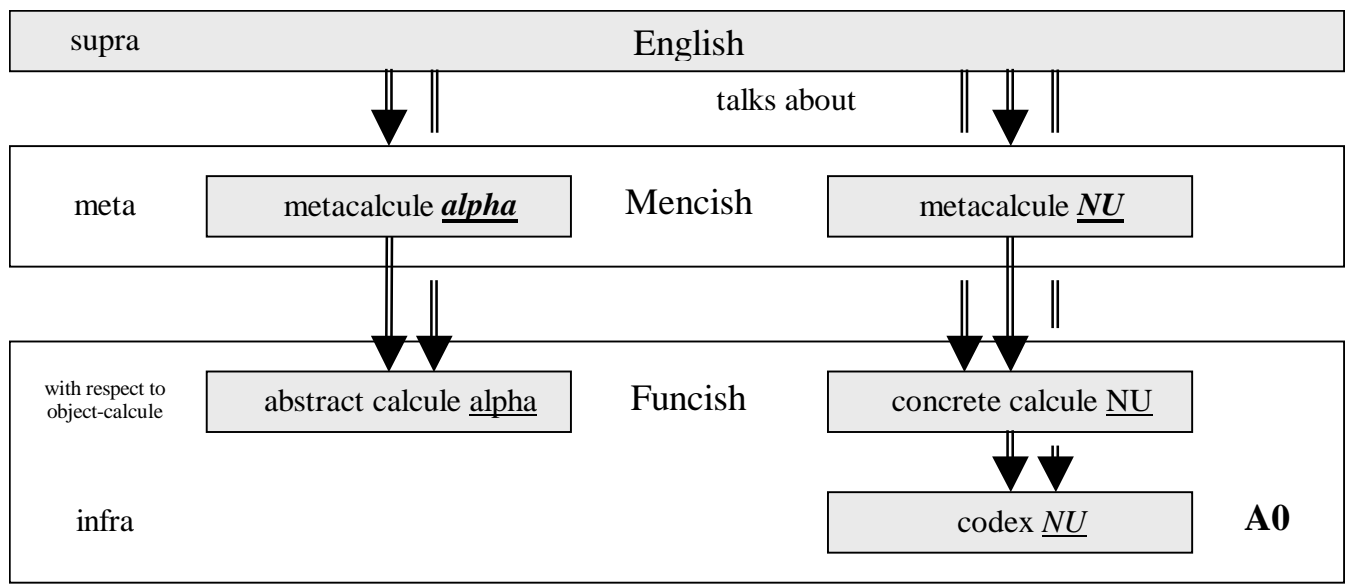

*Figure 1. Hierarchy of languages and codices (tiers of languages)*

A hierarchy of languages means that languages appear in **tiers**\*) : with respect to a given language the next higher tiers are called meta-tier, meta-meta-tier and so on, the next lower tier is called infra-tier. The highest tier is called supra-tier (it is usually the common language), the lowest is called hypo-tier. Two languages with a common metalanguage share the same tier. In this publication the supra-tier is the meta-meta-tier and the infra-tier is the same as the hypo-tier. In this publication I will do some things in English that should be properly done in Mencish. This is for shortness and easier readability and is to be taken care of in future publications.

Mencish is in a sense simpler than the languages it talks about. It talks about finite strings of characters, which means that it is something like a concrete arithmetic calcule, which also talks about finite strings of characters, that are called numbers. Of course, it is not inherent in numbers that they have to be written decimal form. Unal, dual, octal in general multal\*\*), you can write numbers to any base. You may build numbers from the characters of the calcule that a metacalcule is talking about. In this sense all the **metaindividuals**\*\*), i.e. strings, are numbers. I will make ample use of this simple fact. Finally: the two calcules of this publication and their metacalcules are strictly first-order-logic .

## 2.3 Bavaria-notation

This is the idea of bringing good order and maybe even some beauty:

| (idea 3) | **Bavaria-notation** with typographic distinction between languages. |
|---|---|

For the relatively simple cases that are treated in this publication the notation for Mencish and Funcish looks very similar to the usual logical notation. Just for kicks I call it **Bavaria-notation**[**)]. It is computer-proof, you must not change the style of a single character. The three languages English, Mencish and Funcish each have their own alphabet, so that they can already be distinguished by their typography.

The individuals sorts of abstract calcules are denoted by small Greek letters point 12, individual sorts of concrete calcules are denoted by capital Greek letters point 12, e.g. for my two calcules I have $\alpha$ and N. The metaindividual sorts (strings of the corresponding metacalcules) are in boldface italics $\boldsymbol{\alpha}$ and $\boldsymbol{N}$.

Bavaria-notation obeys rule the that you can tell from the name of an entity its exact ontological placement in the system, or in simple language: the names of entities speak. The **name** of the binary multiplication function in abstract calcule alpha is e.g. $\alpha\times(\alpha;\alpha)$ which shows that it is binary and that it maps the two numbers of the argument to a number. Notice that the name is **not** $\alpha\times$ , which in Funcish would be a number-constant like nullum $\alpha$n .

No more further theory for the moment, let me specify the alphabets I need and give an example.

| font Times Roman in various points and styles: |
|---|
| a b c d e f g h i j k l m n o p q r s t u v w x y z |
| A B C D E F G H I J K L M N O P Q R S T U V W X Y Z |
| 0 1 2 3 4 5 6 7 8 9 , . ; . ! ? " ( ) [ ] é ä ö ü ß |
| ' + - * / =                                    # & § $ € @ |

*Table 1. Alphabet for common language English (as you have already noticed in this publication)*

| font Symbol boldface italics | point[1] |
|---|---|
| $\boldsymbol{\alpha}$  $\boldsymbol{N}$  $=$  $\neq$  $\neg$  $\vee$  $\wedge$  $\rightarrow$  $\leftrightarrow$  $($  $;$  $)$  $\exists$  $\forall$  $[$  $]$ | 12 |
| font Arial boldface italics | |
| *0 1 2 3 4 5 6 7 8 9* | 8 |
| *a b c d e f g h i j k l m n o p q r s t u v w x y z* - | 10 |
| *A B C D E F G H I J K L M N O P Q R S T U V W X Y Z* _ | 10 |
| *: ; \* | 12 |

[1] specifications in points apply only if the manuscript is printed in original DIN A4 size

*Table 2. Alphabet of Mencish metacalcules **alpha** and **NU** relating to calcules alpha and NU resp.*

| font Symbol | point |
|---|---|
| $\alpha$  $=$  $\neq$  $\neg$  $\vee$  $\wedge$  $\rightarrow$  $\leftrightarrow$  $($  $;$  $)$  $\exists$  $\forall$  $[$  $]$  $+$  $\times$ | 12 |
| font Arial | |
| 0 1 2 3 4 5 6 7 8 9 | 8 |
| n | 10 |
| ' | 12 |

*Table 3. Alphabet of abstract calcule alpha in Funcish*

Bavaria-notation solves the quotation problem in a perfect fashion: when you talk in one language about the words of another language you just fill them in without danger of mixing up tiers.

## 2.4 Exemplary abstract calcule alpha of arithmetic natural numbers

As an example for the Bavaria-notation and the proper use of language and metalanguage I specify some strings of the first object-calcule**[)], the abstract calcule alpha of arithmetic natural numbers. In short notation I write the following definitions of the metaproperties of strings of alpha (notice the difference between boldface italics of metalanguage and straight letters of calcule language and the use of **concatenation***[)] for strings) :

**nullum ::**           $\alpha n$
meaning              *[ nullum( $\alpha n$ ) ] $\wedge$ [ $\forall \alpha_1$ [ [ $\alpha_1 \neq \alpha n$ ] $\rightarrow$ [ $\neg$ [ nullum( $\alpha_1$ ) ] ] ] ]*

**small-cipher ::**    1 ¦ 2 ¦ 3 ¦ 4 ¦ 5 ¦ 6 ¦ 7 ¦ 8 ¦ 9
**succession ::**      $\alpha'(\alpha)$
**addition ::**        $\alpha+(\alpha;\alpha)$
**multiplication ::**  $\alpha\times(\alpha;\alpha)$
**small-index::**      *small-cipher* ¦ *small-index small-cipher* ¦ *small-index* 0
**number-variable ::** $\alpha$ *small-index*

The nine axioms of the abstract calcule alpha of arithmetic natural numbers are certain strings, where you please notice the subtle difference between boldface italics and normal style of characters ($\alpha_2 \neq \alpha n$ is a true and $\alpha_2 = \alpha n$ is a false metasentence) :

*$\alpha$Aa =* $\forall \alpha_1$ [ $\alpha'(\alpha_1) \neq \alpha n$ ]
*$\alpha$Ab =* $\forall \alpha_1$ [ $\forall \alpha_2$ [ [ $\alpha'(\alpha_1) = \alpha'(\alpha_2)$ ] $\rightarrow$ [ $\alpha_1 = \alpha_2$ ] ] ]
*$\alpha$Ac =* $\forall \alpha_1$ [ $\alpha+(\alpha_1;\alpha n) = \alpha_1$ ]
*$\alpha$Ad =* $\forall \alpha_1$ [ $\forall \alpha_2$ [ $\alpha+(\alpha_1;\alpha'(\alpha_2)) = \alpha'(\alpha+(\alpha_1;\alpha_2))$ ] ]
*$\alpha$Ae =* $\forall \alpha_1$ [ $\alpha\times(\alpha_1;\alpha n) = \alpha n$ ]
*$\alpha$Af =* $\forall \alpha_1$ [ $\forall \alpha_2$ [ $\alpha\times(\alpha_1;\alpha'(\alpha_2)) = \alpha+(\alpha\times(\alpha_1;\alpha_2);\alpha_1)$ ] ]
*$\alpha$Ag =* $\forall \alpha_1$ [ $\forall \alpha_2$ [ [ $\alpha+(\alpha_1;\alpha_2) = \alpha n$ ] $\rightarrow$ [ $\alpha_2 = \alpha n$ ] ] ]
*$\alpha$Ah =* $\forall \alpha_1$ [ $\forall \alpha_2$ [ [ $\exists \alpha_3$ [ $\alpha+(\alpha_1;\alpha'(\alpha_3)) = \alpha_2$ ] ] $\leftrightarrow$ [ [ $\exists \alpha_4$ [ $\alpha+(\alpha_1;\alpha'(\alpha_4)) = \alpha_2$ ] ] $\vee$ [ $\alpha_1 = \alpha_2$ ] ] ] ]
*$\alpha$Ai =* $\forall \alpha_1$ [ $\forall \alpha_2$ [ [ $\exists \alpha_3$ [ [ $\alpha+((\alpha_1;\alpha'(\alpha_3)) = \alpha_2$ ] $\vee$ [ $\alpha+((\alpha_2;\alpha'(\alpha_3)) = \alpha_1$ ] ] ] $\vee$ [ $\alpha_1 = \alpha_2$ ] ] ]

This set of axioms is not **categorical***[)] [7], which means that not all concrete calcules that fulfill these axioms are isomorphic: the correspondences between those concrete calcules are not bijective. This is shown according to Boolos et al. p. 216 [7] : one takes "normal" arithmetics with succession, addition and multiplication, say of decimal numbers as concretisation (I) and constructs the concretisation (II) from (I) by adding to extra numbers with function tables that are extended for these values. Whereas concretisation (I) has commutativity both of addition and multiplication, concretisation (II) has not.

Thus abstract calule alpha of arithmetic natural numbers contains **indefinite***[)] sentences, like e.g.

*[ $\neg$ [ Truth( $\forall \alpha_1$ [ $\forall \alpha_2$ [ $\alpha+(\alpha_1;\alpha_2) = \alpha+(\alpha_2;\alpha_1)$ ] ] )] ] $\wedge$*
*[ $\neg$ [ Falsehood( $\forall \alpha_1$ [ $\forall \alpha_2$ [ $\alpha+(\alpha_1;\alpha_2) = \alpha+(\alpha_2;\alpha_1)$ ] ] )] ]*

Let me point out an important general feature: to every true sentence of a calcule there exist a true metasentence in its metacalcule, namely just the metasentence that states the truth. E.g. the above first axiom is a true sentence of abstract calcule alpha of arithmetic natural numbers: $\forall \alpha_1$ [ $\alpha'(\alpha_1) \neq \alpha n$ ]

The corresonding metasentence is: *Truth( $\forall \alpha_1$ [ $\alpha'(\alpha_1) \neq \alpha n$ ] )*

So far abstract calcules, they will not be investigated any further in this publication.

## 3. Concrete calcule NU of decimal arbation natural numbers

### 3.1 Decimal numbers with a twist

It is just for convention that I present the concrete calcule NU as a calcule for decimal numbers. Instead of ten I might as well use any other base greater than three. Quartal numbers (with 0 1 2 3 ) would do fine, whereas dual numbers (with 0 1 ) would pose some technical troubles; the troubles could be overcome but it is not worth it as the reader will see.

| font Arial | point |
|---|---|
| 0  1  2  3  4  5  6  7  8  9  {  }  ,                                      〈  〉 | 12 |

Table 4. Alphabet for codex NU

| font Symbol | point |
|---|---|
| N[1])  =  ≠  ¬  ∨  ∧  →  ↔  (  ;  )  ∃  ∀  [  ]  +  × | 12 |
| font Arial for concrete calcule NU | |
| 0  1  2  3  4  5  6  7  8  9 | 8 |
| a  b  c  d  e  f  g  h  i  j  k  l  m  n  o  p  q  r  s  t  u  v  w  x  y  z  - | 10 |
| '  ∧ | 12 |

[1]) read this Greek letter "capital nu", don't ever say "en"; watch the difference from Times Roman character N: N

Table 5. Alphabet of concrete calcule NU

Concrete calcule NU of decimal arbation natural numbers talks about the **number** strings of codex *NU* that I write in a funny fashion using three synonymous characters.

| instead of | 8 | I write | { | and read "left brace" or "acco" |
| instead of | 9 | I write | } | and read "right brace" or "lade" |
| instead of | 89 or { } | I write | , | and read "comma" |

The comma is not really part of the language. I use it like a makro of a programming language that is to be expanded whenever a string is processed. It is just for better understanding of **number** strings. A matching pair of characters { ... } I call an "accolade".

| ***acco**\*) ::* | { |
| ***lade**\*) ::* | } |
| ***decimal-cipher ::*** | 1 ¦ 2 ¦ 3 ¦ 4 ¦ 5 ¦ 6 ¦ 7 ¦ { ¦ } |
| ***decimal-numeral ::*** | 0 ¦ decimal-cipher |
| ***positive-number ::*** | decimal-cipher ¦ positive-number  decimal-numeral |
| ***number ::*** | 0 ¦ positive-number |

I count: zero, one, two, three, four, five, six, seven, eight resp. acco, nine resp. lade and write e.g. my year of birth nineteenhundred-and-fortyone 1}41 ; one can get used to that and the reader will see very soon why I do that. I will also use the following **number** strings:

| ***octal-cipher ::*** | 1 ¦ 2 ¦ 3 ¦ 4 ¦ 5 ¦ 6 ¦ 7 |
| ***octal-numeral ::*** | 0 ¦ octal-cipher |
| ***positive-octal-number ::*** | octal-cipher ¦ positive-octal-number  octal-numeral |
| ***octal-number ::*** | 0 ¦ positive- octal-number |
| ***field ::*** | octal-number ¦ 0 positive-octal-number |

## 3.2 The decimal arbation calculator ARBATOR

Before I start talking about calculators lets have some definitions:

**small-letter-symbol ::**     a ¦ b ¦ c ¦ ... ¦ x ¦ y ¦ z ¦ - ¦ ' ¦ + ¦ × ¦ ^
**small-word ::**              *small-letter-symbol* ¦ *small-word small-letter-symbol*
**number-constant ::**         N *small-word*

**number-constant** strings are names of **number** strings, e.g. for **nullum** I have **number-constant(**Nn**)**

**zero ::**                    0
**nullum ::**                   Nn
**nullum-thing**\*[)] **::**    *zero* ¦ *nullum*

**number-variable ::**         N *small-index*
**number-dingus**\*[)] **::**   *number* ¦ *number-constant* ¦ *number-variable*
**number-thing ::**            *number-variable* ¦ *number*
**number-array ::**            *number* ¦ *number-array* ; *number*
**number-argument ::**         ( ) ¦ ( *number-array* )
**number-variable-array ::**   *number-variable* ¦ *number-variable-array* ; *number-variable*
**number-variable-argument ::** ( ) ¦ ( *number-variable-array* )
**number-dingus-array ::**     *number-dingus* ¦ *number-dingus-array* ; *number-dingus*
**number-dingus-argument ::**  ( ) ¦ ( *number-dingus-array* )

So far the codex <u>NU</u> that concrete calcule <u>NU</u> is talking about has only **number** strings. In order to do some mathematics I need some mappings or relations. As it was introduced in section 1.1. a codex can contain calculators, that take care of functions and relations. Codex <u>NU</u> contains a series of functions that I call the decimal **arbation** calculator or decimal **ARBATOR**\*\*[)] which is an acronym for ARBation calculATOR:

Narba(N)
Narba(N;N)
Narba(N;N;N)
...

and so on, where I call the first argument position the **program**-position\*[)] and the consecutive the **input-**positions\*[)]. As I do not have any other functions I will use a synonymous notation with the prodecure number behind the argument without the danger of any confusion:

$\forall N_1$ [ Narba($N_1$) = ( )$N_1$ ]
$\forall N_1$ [ $\forall N_2$ [ Narba($N_1;N_2$) = ($N_2$)$N_1$ ] ]
$\forall N_1$ [ $\forall N_2$ [ $\forall N_3$ [ Narba($N_1;N_2;N_3$) = ($N_2;N_3$)$N_1$ ] ] ]
...

and so on. This is part of the conventions that I call Bavaria-notation (as opposed to the Polish notation where parentheses are lacking). This abbreviation is only used in concrete calcule <u>NU</u> .

The decimal arbation calculators that I am going to define in the next two chapters produce exactly the usual **primitive recursive** functions. Decimal arbation calculators are just another method to calculate primitive recursive functions. However, it lends itself to some new ideas that will turn out to be very useful. Decimal arbation calculators make use of the decimal[1] **ARBACUS**\*\*[)] computer that I am going to describe in the following. ARBACUS is an acronym for ARBation abaCUS.

---

[1] Until section 7.5 I will only talk about **decimal** ARBACUS and ARBATOR and therefore will leave away **decimal**.

## 3.3 Definition of arbor-number and primitive-arbor-number strings

Firstly I define two important classes of numbers, which will be the basis of arbation.

In the following **f** is mnemonic for **field** and **a** for **accolade**\*), denoting whether a **number** string starts or ends with a **field** or an **accolade** respectively.

| | |
|---|---|
| **f-f-tree ::** | field ¦ field a-f-tree ¦ f-a-tree field ¦ field a-a-tree field |
| **accolade-tree ::** | {} ¦ { f-f-tree } |
| **a-a-tree ::** | accolade-tree ¦ a-a-tree a-a-tree ¦ a-a-tree f-a-tree ¦ a-f-tree a-a-tree |
| **f-a-tree ::** | f-f-tree a-a-tree |
| **a-f-tree ::** | a-a-tree f-f-tree |
| **arbor-number ::** | octal-number ¦ positive-octal-number a-f-tree ¦ |
| | positive-octal-number a-a-tree ¦ a-f-tree ¦ a-a-tree |

Now one can see why I have chosen the word **arbor** : it is Latin for tree. And this leeds to ARBACUS, ARBATOR, arbation etc.

For short: an **arbor-number**\*) string has a tree-structure through matching characters { and } , where no {{ or }} are admissible, furthermore it does not contain multiply prenulled octal numbers.

E.g.{1 {0} 0,01} is an **arbor-number** string, 10{{ and {001} are not **arbor-number** strings.

A **number** string that is not an **arbor-number** string is called a **herbum-number** string.

A branch of a tree is called **accolade**; an **accolade** starts with an **acco** and finishes with a **lade**. The first field of an accolade is called its **counter**\*), the last field its **limit**\*).

A **primitive-arbor-number**\*) string is an **arbor-number** string where counter and limit fields of all **accolade** do only appear inside the accolade as limit fields or in the fast-finish-form\*) { counter , limit } .

E.g. {1{1}2} is not a **primitive-arbor-number** string as the counter 1 appears within the **accolade**, {0{0,01}01} is a **primitive-arbor-number** string altough the counter appears within the **accolade** but in the admissible form.

The meaning of **primitive-arbor-number** strings will become clear soon: they are the **number** strings that lead exactly to the primitive recursive functions as they are known in normal recursive function theory [6] . If one uses them in the program field for a binary function say of addition Narba(N+;N1;N2) = (N1;N2)N+ it is guaranteed that the computation halts for all input.

An **arbor-number** string that is not a **primitive-arbor-number** string is called a **complex-arbor-number**\*) string.

In the arithmetical universum of **number** strings **arbor-number** strings are very scarce and even more so **primitive-arbor-number** strings, you may compare it to stars in the relatively empty physical universe. And yet, what a beautiful and big world is the physical universe and what a beautiful and big world is the arithmetical universe!

## *4. ARBACUS computer*

### *4.1 Direct coding and fields*

I always found Gödel numbering something weird, as it is entering the considered systems from the **outside**. That the original method used prime numbers had its special yet strange touch but did not bother me. That and my deep trust in the finity of language is what lead me to

| (idea 4) | **direct coding** instead of cumbersome Gödelisation. |
|---|---|

By **direct coding** I mean that every *number* string can be interpreted as a **primitive recursive** function and that for every **primitive recursive** function there is a *number* string which is its code. Codes and *number* strings are the same. You will get a first feeling what I mean by a coding of functions by numbers if you look at the following example, where you will also understand why I write the numbers 8 and 9 with synonymous characters { and } , which leeds to a tree structure.

The binary function of addition x+y of two numbers is coded by the *number* string 808901981890890 or synonymous form {0,01}{1,0,02} and given the name M+ as *number-constant*. It means that there are input fields 01 and 02 , that the value ⟨01⟩ of the field 01 is put into field 0 and then the value of this field 0 is incremented by one when the scratch field 1 runs from value 1 to the value ⟨02⟩ of field 02 , if the field 02 has value 0 nothing is done in the second part. As you see accolades { ... } are used as **loops***⁾ (like do-loops in applied computing).

Like an Abacus or a Register computer the ARBACUS computer has an unlimited **memory** with unlimited **fields***⁾ that are used during the computation to store **values**, e.g. ⟨01⟩ as value field 01 , where the symbols ⟨ ⟩ are just used inside the codex. There is an additional **register***⁾ field in the memory that is not addressable inside a program, it contains the *number* of the program to be computed. The fields are referenced by numbers according to the following convention:

| | | | | | | | | | | program-register field | 00 | | | | | | | | | | |
|---|---|---|---|---|---|---|---|---|---|---|---|---|---|---|---|---|---|---|---|---|---|
| | | input fields by prenulled octal numbers | | | | | | | | | | output field | | | scratch fields by octal numbers | | | | | | |
| ... | 012 | 011 | 010 | 07 | 06 | 05 | 04 | 03 | 02 | 01 | | 0 | 1 | 2 | 3 | 4 | 5 | 6 | 7 | 10 | 11 | 12 | ... |

*Table 6. ARBACUS computer memory fields*

**output-field***⁾ *::*          0
**input-field***⁾ *::*           0 *octal-cipher* ¦ *input-field  octal-numeral*
**scratch-field***⁾*::*        *octal-cipher* ¦ *scratch-field  octal-numeral*
**field ::**                      *output-field* ¦ *input-field* ¦ *scratch-field*
**program-register ::**   00

| | | | description with fields | | | description with values | |
|---|---|---|---|---|---|---|---|
| *prog. reg. f.* | *input fields* | | 00 | 01, 02, ... | | *prog. value* | *input values* |
| arity Arbacus computation | | | arity Arbacus computation | | | arity Arbacus computation | |
| *output field* | | | 0 | | | ⟨0⟩ | |

*Figure 2. ARBACUS computer with fields*

A computer works step by step. By **step** I mean the smallest units in a computation, that will be given in the next section for the ARBACUS computer.

*4.2 Computation rules*

After these preliminaries I have to go into the details of

> (idea 5) **ARBACUS computer** with a new programming language without referenced branching.

I will define the new programming language that I call **A0**. It is a command language as opposed to proposition languages Funcish and Mencish. Actually it is an interpretative language as will be seen (in applied computing the best known interpretative language is BASIC). A program is given by an *arbor-number* string, which is interpreted in the following way. Besides the **empty** command {} that is abbreviated by the comma there are only the two following **commands**:

S     **succeed**, replace the actual value of a field by its successor e.g.    12

R     **repeat** performing an accolade {...} of commands       e.g.    {1 , 2{4} 3}
       enclosed by acco and lade a certain number of times,
       as given by the limit value in the lade-field with the acco-field carrying the counter of the step,
       do nothing if the limit value is zero, in this case the counter field has value zero after performing;
       before the accolade is performed the acco-field is set to zero
       at the end the acco-field contains the the limit value of the lade-field

Two special cases of the repeat command: only one field or one pair of fields in accolade:

D     **delete,** put the value to zero                         e.g.    {13}

C     **copy** the value of the second field to the first           e.g.    {4,03}

At the start of a computation firstly all fields are **initialised** to zero, then the input fields 01  02  03 ... to the values of the input argument starting from the left. If the arity of the argument is higher than the highest input field just ignore the higher ones. If the arity of the argument is less than the highest input field the values of the exceeding input fields are put to zero through the initialisation.

**Computation** starts from the left and proceeds to the right, a cursor moves through the digits of the *number* string. The output is in field 0. The only backspacing can occur at the end of an accolade. This is where non-halting may occur, e.g. *arbor-number* 1{0{0}1} never halts, as you see from the table:

| step | field 0 | field 1 | command |
|------|---------|---------|---------|
| 0 | 0 | 0 | initialise |
| 1 | 0 | 1 | succeed |
| 2 | 0 | 1 | initial value |
| 3 | 1 | 1 | succeed as limit not yet reached |
| 4 | 0 | 1 | delete |
| 5 | 1 | 1 | succeed as limit not yet reached |
| 6 | 0 | 1 | delete |
| ... | | | and so on forever |

*Table 7. example program computation*

The above example is not a *primitive-arbor-number* string. All *primitive-arbor-number* programs halt when applied to any input. The beauty of programming language **A0** lies in the fact that one can check the sufficient condition of program halt that governs all primitive recursive functions. It is trivial that all primitive recursive functions and only the primitive recursive functions can be obtained by applying the ARBACUS computer to a *primitive-arbor-number* string as program. The power of **A0**: you can program " if-then-else" (let  Na and Nb be two example numbers without scratch-field collision) :

if field 1 has value 0 do Na else do Nb :         {3}3 {2 Nb {3} {2,1} 1} {2 Na 3}

## 4.3 Examples of primitive-arbor-number strings

| number-constant | description | number | arity |
|---|---|---|---|
| Nn | unification**) | 0 | 0 |
| Nnf | nullification**), nullum constantion**) [1] | {0} | 0 |
| Ndf | duofication**), | {0}0,0 | 0 |
| Ndef | decification**), | {0}0,0,0,0,0,0,0,0,0 | 0 |
| Nid | identity, identation**) | {0,01} | 1 |
| Ndpj | bi-projection | {0,02} | 2 |
| Ntpj | tri-projection | {0,03} | 3 |
| N' | succession | {0,01} 0 | 1 |
| Nsig | signum, signation**), | {1 {0} 0,01} | 1 |
| Nneg | not, logical negation | {0} 0 {1 {0} 01} | 1 |
| Nand | and, logical addition | {1,01} {2,1,02} {2 {0} 0,1} | 2 |
| Nlor | or, logical multiplication | {3} {2 {1,3,01} 02} {2 {0} 0,3} | 2 |
| N+ | addition x+y | {0,01}{1,0,02} | 2 |
| N× | multiplication x*y | {2 {1,0,01} 02} | 2 |
| Nxp | exponentiation $x^y$ | {0} 0 {1 {2,0} {0} {3 {4,0,01} 2} 02} | 2 |
| Nsuxp | superexponentiation | {0}0{7{4,0}{0}0{5{2,0}{0}{3{1,0,01}2}4}02} | 2 |
| Nfac | factorial x! | {0} 0 {4 {1,0} {0} {2 {3,0,1} 4} 01} | 1 |
| Nprd | predecession | {2} {1 {0,2} 2,01} | 1 |
| Ntst | truncated subtraction [x-y] | {0,01} {1 {2,0} {3} {4{0,3} 3,2} 02} | 2 |
| Nadi | absolute difference | {5,01} {1 {2,5} {3} {4{5,3} 3,2} 02} {0,02} {1 {2,0} {3} {4{0,3} 3,2} 01} {1,0,5} | 2 |
| Nevy | evenness characteristic[2] | {1 {3,0} {0}0 {2 {0} 3 } 01} | 1 |
| Nody | oddity characteristic | {0} 0 {1 {3,0} {0}0 {2 {0} 3 } 01} | 1 |
| Ndiv | entire division [x/y], if divide by zero successor | {6,01} 6 {7,6} {5 {1 {2,6} {3} {4{6,3} 3,2} 02} {1 {0,5} 6} 7} | 2 |
| Ndir | entire division remainder, x-[x/y]*y if divide by zero identity | {0,01} 0 {7,0} {5 {1 {2,0} {3} {4{0,3} 3,2} 02} {1 {6,0} 0} 7} {2} {1 {0,2} 2,6} | 2 |
| Nrt | entire root [$^x$root y] is zero if y zero is y if x zero | {6 {7} 7 {1 {2,7} {7} 3 {4,7,6} 2} 01} {5,02} 5 {1 {2,5} {3} {4{5,3} 3,2} 7} {2} {1 {2} 2,5} {1,0,2} 02} | 2 |
| Nlg | entire logarithm [log$_x$y] is zero if y zero is y if x zero | {0}0 {7} {6 {3,0} {2 {4,0,01} 3} 7 {1 {2,02} 2 {3} {4 {5,3} 3,2} 0} {1} {2 {1} 1,5} {2 {6,02} 1} 02} {0} {2 {1 {0,2} 2,7} | 2 |
| Neqy | equality characteristic | {5,01} {1 {2,5} {3} {4{5,3} 3,2} 02} {0,02} {1 {2,0} {3} {4{0,3} 3,2} 01} {1,0,5} {1,0} {0} {2 {0} 0,1} | 2 |
| Niey | inequality characteristic | {5,01} {1 {2,5} {3} {4{5,3} 3,2} 02} {0,02} {1 {2,0} {3} {4{0,3} 3,2} 01} {1,0,5} {1,0} {0} 0 {2 {0} 1} | 2 |
| Nminy | minority characteristic | {2,01} 2 {3} {4 {1,3} 3,2} 02} {0} {2{0} 0,1} | 2 |
| Nemiy | equal-minority charact. | {2,01} {3} {4 {1,3} 3,2} 02} {0} {2{0} 0,1} | 2 |

[1] constantion: it gives constant value  [2] warning: in characteristic functions I choose truth value 0 and falsity value 1.
I think it is nicer to represent the logical "and" by "plus"; in applied computing one has the error code zero for "no error".

*Table 8. Some important **primitive-arbor-number** strings (to be continued)*

| number-constant | description | number | arity |
|---|---|---|---|
| Npry | primality<br>2 3 5 ... | {0}0 {11} {1} {2 {11,1} 1,01} 1<br>{2 {7 {6 {3,10} {4} {5{10,4} 4,3} 2}<br>{3 {0,10} 10} 1}<br>{3} {4} {5 {3,4} 4,0} {0}0 {5{0}3} {3{2,11}0} 11} | 1 |
| Nnpr | non-primality<br>0 1 4 6 ... | {{0} {11} {1} {2 {11,1} 1,01} 1<br>{2 {7 {6 {3,10} {4} {5{10,4} 4,3} 2}<br>{3 {0,10} 10} 1}<br>{3} {4} {5 {3,4} 4,0} {0}0 {5 {0} {2,11} 3} 11} | 1 |
|  | *as exercises* | *comment* |  |
| Ndecc | decimal **concatenation** e.g.<br>{0}0 and {7} gives {0}0{7} | I write concatenation in short Bavaria-notation[1]:<br>(N1;N2)Ndecc = N1 N2 | 2 |
| Nprn | prime denumeration<br>0 for 0, 2 for 1, 3 for 2 ...<br>using majorant Nsuxp |  | 1 |
| Npair | pair antidiagonal method<br>$((j+k)^2+3j+k)/2$ | bijectively coding a pair of numbers by one number | 2 |
| Nrow | row antidiagonal method<br>$n-(d(n)(d(n)-1))/2$<br>where $d(n) =$<br>$[(1+ \text{entire } ^2\text{root}(1+8n))/2]$ | (see table below) | 1 |
| Ncol | column antidiagonal method<br>$(d(n)(d(n)+1))/2-1-n$ |  | 1 |
|  | *see chapter 5* |  |  |
| Narby | arbor-number characteristic |  | 1 |
| Nupary | unary arbor-number characteristic |  | 1 |
| Ndpary | binary arbor-number characteristic |  | 1 |
| ... |  |  |  |
| Nhery | herbum-number characteristic |  | 1 |
| Npary | primitive-arbor-number characteristic |  | 1 |
| Ncary | complex-arbor-number characteristic |  | 1 |
|  | *see chapter 6* |  |  |
| Nhxpg | hyperexponentiation generator |  | 1 |
| Nhicg | hyperincrementation generator |  | 1 |

|  |  | columns k |  |  |  |  |  |  |  |
|---|---|---|---|---|---|---|---|---|---|
| rows j | 0 | 1 | 2 | 3 | 4 | 5 | 6 | 7 | ... |
| 0 | 0 | 1 | 3 | 6 | 10 | 15 | 21 | 2{ |  |
| 1 | 2 | 4 | 7 | 11 | 16 | 22 | 2} |  |  |
| 2 | 5 | { | 12 | 17 | 23 | 30 |  |  |  |
| 3 | } | 13 | 1{ | 24 | 31 |  |  |  |  |
| 4 | 14 | 1} | 25 | 32 |  |  |  |  |  |
| 5 | 20 | 26 | 33 |  |  |  |  |  |  |
| 6 | 27 | 34 |  |  |  |  |  |  |  |
| 7 | 35 |  |  |  |  |  |  |  |  |
| ... |  |  |  |  |  |  |  |  |  |

[1] does not collide with other string forming rules

*Table 8. Some important **primitive-arbor-number** strings (continuation)*

I rewrite the **number** Niey for the equality characteristic of the above table. With the usual characters for eight and nine I get the following more familiar form of a number that is close to $10^{95}$. If you think that this is a big number, just wait for sections 6.5 and 6.6 :

8589019818 2895983984 8589393892 9029808902 9818289098 3984808939 3892901981
8908959818 9098092808 0919

## 5. ARBATOR *calculator*

### *5.1 Bootstrap mechanism*

So far I have described the action of an ARBACUS computer when it is fed a **primitive-arbor-number** as program. What happens in the two cases when the given **number** is not a **primitive-arbor-number** ?

In the first case it is a **herbum-number** , i.e. not an **arbor-number** . The above rules for computation cannot be applied and therefore I assign arbitrarily that the ARBACUS computer does not halt in this case, but rather keeps on forever.

In the second case it is a **complex-arbor-number**. In this case one can apply the above rules but there are two possibilities when applied to a certain input, either the computer halts after a finite number of steps or it does not. At the moment I do not go into the question if it can be decided whether it halts or not. In any case that is where the problems may arise, the predetermined breaking point, if you wish.

The important thing is that there exist **primitive-arbor-number** strings Narby, Nhery, Npary and Ncary that give rise to characteristic functions via arbation by which it can be checked if a **number** string is

- **arbor-number**
- **herbum-number**$^{*)}$
- **primitive-arbor-number**
- **complex-arbor-number**

$\forall N_1 \ [ \ [ \ \text{arbor-number}(N_1) \ ] \leftrightarrow [ \ \text{Truth}( \ (N_1)\text{Narby} = 0 \ ) \ ] \ ]$
$\forall N_1 \ [ \ [ \ \text{herbum-number}(N_1) \ ] \leftrightarrow [ \ \text{Truth}( \ (N_1)\text{Nhery} = 0 \ ) \ ] \ ]$
$\forall N_1 \ [ \ [ \ \text{primitive-arbor-number}(N_1) \ ] \leftrightarrow [ \ \text{Truth}( \ (N_1)\text{Npary} = 0 \ ) \ ] \ ]$
$\forall N_1 \ [ \ [ \ \text{complex-arbor-number}(N_1) \ ] \leftrightarrow [ \ \text{Truth}( \ (N_1)\text{Nxary} = 0 \ ) \ ] \ ]$

I do not go into the concept of truth in this publication, just note that I have written the metaproperty **Truth** with a first capital letter; by this I indicate that this is in general not a decidable metaproperty, whereas a metaproperty like **arbor-number** is decidable.

I talk about a **bootstrap**$^{*)}$ mechanism as one can apply Narby to itself and gets (Narby)Narby = 0 thereby stating **primitive-arbor-number(** Narby **)**

These **primitive-arbor-number** strings are rather difficult to construct and are not developped in this publication. I just sketch how a programmer has to proceed in the construction of Narby . If you are familiar with primitive recursive functions it is immediately clear that these characteristic functions are primitive recursive. The necessary loops that run over all characters from left to right the **number** string have of a luxurious majorant given by the **number** string itself.

- Firstly one has to check the acco-lade-structure: starting from the left one checks digit by digit if the count of accos { never gets below the count of lades } and that if at the right end the two counts match.
- Secondly one checks that no {{ or }} occur
- Thirdly one checks that it does not contain multiply prenulled octal numbers.

For the construction of Npary one fourthly checks that the *counter* and *limit* fields of all accolades do at most appear inside the accolade as limit fields or in the fast-finish-form { *counter* , *limit* } .

The intrinsic top-down-structure of the command language **A0** allows for a check if a **number** string is a **primitive-arbor-number** string. I do not see a similar possibility for a command language with referenced branchings (like in Abacus- or Register-programs).

## 5.2 Two-layer-computation for arbation

Now everything is prepared to define the full action of the **calculators** Narba(N;...;N) when applied to arguments of **program** in first position and normal input in the adjoining positions, e.g. multiplication N× with binary input:

Narba( {{2{1,0,01}02} ;734;1{ ) = (734;1{ ) {2{1,0,01}02}

**primitive-term**[*] **::**      **number-argument number**

Every **number** can be applied to every **number-argument**. If the arity of the **number** does not coincide with the arity of the **number-argument** the calculators obeys the following rule:

- argument positions that are higher than the arity of the **number** are ignored
- missing argument positions that are required due to the arity of the **number** are taken as **zero**

Remember, per definitionem a computer may or may not halt, a calculator always halts. A program can be performed on a computer or on a calculator. Now I am going to construct the arbation calculator from the ARBACUS computer. To this end I introduce

(idea 6) **ARBATOR two-layer-calculator** for primitive recursion, as a new paradigm.

In a single calculation more than one application of a computer may occur; infinitely many applications would not make sense. Some finite logic may connect the various **layers**[*]. The important rule for the multiple application of computers within calculators is that no-halt-situations are excluded. There is no reason why the so defined calculator should not be used more than once, i.e. in many **levels**[*] of a calculation, as I will show in the next section.

In concrete calcule NU **two layers** are sufficient, where I use bootstrap mechanism with Npary on layer 1 in order to check if the given program is a **primitive-arbor-number** string; you get this result after a finite count of steps. On layer 2 the actual computation is performed; as no non-halting loops can occur you get a result after a finite count of steps. For **herbum-number** strings the trivial result is **zero**, I call them **Nully** [**]. The overwhelming majority of **number** strings is **Nully** (see also remark at the end of section 3.3). The following simple diagram describes the new calculator:

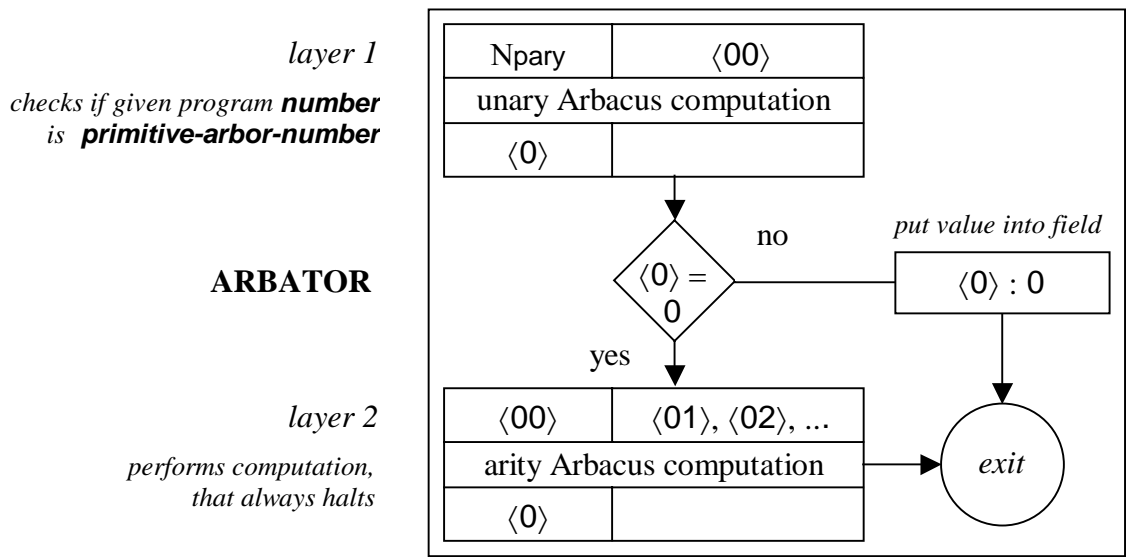

*Figure 3. Flow-diagram arbation calculator (see table 6 and figure 2)*

## 6. Investigating arbations

### 6.1 Definition of *pattern*, *term* and *scheme* strings

In the following I will no longer talk of functions but only of **scheme** strings. As you may have noticed I have only used the word "function" in common language but I have never used a word **function** in metalanguage Mencish. There is deeper meaning in that as functions are equivalence classes of **scheme** strings (of denumerably infinite cardinality), but I will not go into this any further.

I define **pattern**[*)] strings:

| | |
|---|---|
| **pattern-array ::** | **pattern ¦ pattern-array ; pattern** |
| **pattern-argument ::** | **( ) ¦ ( pattern-array )** |
| **pattern ::** | **number-dingus ¦ pattern-argument pattern** |

A **pattern** string without a **number-variable** is called **term**[*)] string, a **term** string can be called a **nullary-pattern**[**)] string. A **pattern** string with at least one **number-variable** is called **scheme**[*)] string. According to the highest appearing **number-variable** one has **positary-scheme**[**)] i.e. **unary-scheme**, **binary-scheme**, **trinary-scheme** ... strings, I speek of free arity.

According to the count of different **number** or **number-constant** strings one has **once-parametric-scheme**, **twice-parametric-scheme**, **thrice-parametric-scheme** ... strings, I speek of parametric arity.

### 6.2 Primitive recursion and primitive-scheme strings

Now that the proper language is installed I am prepared to investigate the concrete calcule <u>NU</u> that gives rise to **arbative**[**)] functions that are given through **scheme** strings and whose evaluation are written as **term** strings.

The calculators of all arities has been designed to perform calculations for a given program **number** string and input **number** strings of given arity, e.g. the addition of 3 and 4 by (3;4){0,01}{1,0,02} = 7 that one can abbreviate by using the **number-constant** string N+ to give (3;4)N+ = 7 . In the language of the preceding section (3;4){0,01}{1,0,02} and (3;4)N+ are **term** strings.

What I am interested in now are **primitive-scheme** strings, e.g. the binary addition with two **number-variable** strings ($N_1$;$N_2$)N+ which I call a **binary-primitive-scheme** string, as it contains **number-variable** strings in the input argument positions and a **number-constant** string in the program position.

| | |
|---|---|
| **primitive-scheme**[*)] **::** | **number-dingus-argument number-constant ¦** |
| | **number-dingus-argument number** |

A **scheme** string that is not a **primitive-scheme** string is called a **complex-scheme**[*)] string.

In section 4.2 it was remarked that it is trivial that all primitive recursive functions and only those can be obtained by applying the ARBACUS computer to a **primitive-arbor-number** string as program. It is also immediately clear that the ARBATOR calculator produces all primitive recursive functions and only those, by applying it to **number** strings; in the first layer all **non-primitive-arbor-number** strings are singled out to produce nullification (function that is always zero). In the second layer the actual calculation is performed. **Primitive arbative** is the same as **primitive recursive**.

One does not have to worry that so many numbers give rise to nullification and that for every function there is an infinity of possible **primitive-arbor-number** strings. There is room enough for everybody. Such is the world of numbers: very big.

## 6.3 Multilevel-calculations for arbation, primative-scheme and exotic-scheme strings

So far it may look that arbation is just another method to calculate primitive recursive functions. But let us look at composition of functions. Composition of functions means that one inserts one function into another, e.g. the simplest case in traditional notation f(g(x)) . Of course one can do this with *scheme* strings and obtain other *scheme* strings, actually I have already done this when recursively defining *scheme* strings in section 6.1 .

In section 3.2 I have already singled out the first argument position of the calculator functions e.g. Narba($N_1$;$N_2$;$N_3$) = ($N_2$;$N_3$)$N_1$ by using synonymous Bavaria-notation, that puts the program number behind the argument. The reason[1] was that the program number (in the example $N_1$) of the first position is treated completely different from the input-argument (in the example $N_2$;$N_3$ ) that follows it. In a completely natural way there appears

| (idea 7) | **multi-level-calculation** including **procession**[*] for non-primitive functions. |
|---|---|

It means that there are two types of compositions, with far reaching consequences. I call it **primative**[**] when the insertion happens in the input-argument and I call it **processive**[**] when the insertion takes place in the program number. If one inserts functions into each other one has various **levels** of composition, that is why I call it **multilevel-calculation**[*].

A *scheme* string is called a *primative-scheme* string, when all insertions are primative, otherwise it is called a *processive-scheme* string. A *scheme* string is called an *orthodox-scheme*[*] string, when no *number-variable* strings appear in the place of program number, otherwise it is called a *paradox-scheme*[*] string. A *scheme* string is called a *conventional-scheme*[*] string, when it is both a *primative-scheme* and an *orthodox-scheme* string. A *scheme* string that is not a *conventional-scheme* string is called an *exotic-scheme*[*] string.

As long as one has *conventional-scheme* strings and inserts them into each other in input positions, or as I say uses **primation**[**] only, one stays in the world of *conventional-scheme* strings, is it closed under this type of composition. This corresponds to the fact that primitive recursive functions are closed under composition.

Theorem A:  $\forall N_1 [ \forall N_2 [ \exists N_3 [ \forall N_4 [((N_4)N_2)N_1 = (N_4)N_3 ] ] ] ]$

Proof idea:
There is a *primitive-arbor-number* Nuuc such that $N_3 = (N_1;N_2)$Nuuc , the result is a concatenation $N_1$ {1}{2}...{01,0} {0} $N_2$ with sufficient scratch field deletions in between.

All theorems of that type can be combined in metatheorems

Metatheorems *A*: Every *conventional-scheme* string can be replaced by a *primitive-scheme* string

$\forall N_1 [ [ \textit{unary-conventional-scheme}(N_1) ] \rightarrow$
$[ \exists N_2 [ [ \textit{number}(N_2) ] \wedge [ \textit{Truth}( \forall N_1 [ \textit{N}_1 = (N_1)\textit{N}_2 ] ) ] ] ]$

$\forall N_1 [ [ \textit{binary-conventional-scheme}(N_1) ] \rightarrow$
$[ \exists N_2 [ [ \textit{number}(N_2) ] \wedge [ \textit{Truth}( \forall N_1 [ \forall N_2 [ \textit{N}_1 = (N_1;N_2)\textit{N}_2 ] ] ) ] ] ]$

...

---

[1] besides abbreviation an aesthetical reason for this notation is given by the direction that the calculator works: first the memory is filled with the values of input and then (after the program check) the computation cursor starts moving from left to right (except for backward jumps at the end of accolades)

## 6.4 Procession and generators

I brought up idea 4 of direct coding which says that every **number** string is a program although the most **number** strings just lead to the constantion zero. But it brings a a completely new quality into calculation: you calculate a **number** string and use this as a new program. When I write it in Bavaria-notation it becomes clear why I call it **procession**, look e.g. at the following **scheme** as an example:
($N_2$)($N_1$){1}1

Like in most cases the first result is trivial, i.e. ($N_1$){1}1 = 0 and as ($N_2$)0 = 1 the result is the unication (it gives constant value 1) .

But I can force interesting results when the first result is itself a nontrivial **primitive-arbor-number** string. Look at the following examples of identation and hidden addition:

($N_2$)($N_1$){0,01} = ($N_2$)$N_1$

($N_1$;$N_2$)({0,01}{1,0,02{)N' = ($N_1$;$N_2$){0,01}{1,0,02} = ($N_1$;$N_2$)N+

Besides similar rather amusing constellations there is the very important case that the first result is a nontrivial **primitive-arbor-number** string for all input. I say that the the first program **primitive-arbor-number** string is a **Generator-number**[*] string. It is not decidable by a general method if a **primitive-arbor-number** string is a **Generator-number** string, but this poses no problem. One has to demonstrate this in every single case.

By the way generator-technique is well established in applied computing (so-called fourth-generation-languages, where here the word "generation" has nothing to do with generating, but with progress).

In section 6.5 and 6.6 I will give important examples for the generator-technique. It will turn out that problems that have lead to the extension from primitive recursive functions to recursive functions (via the inclusion of minimisation) are so much easier solved with **processive-scheme** strings.

## 6.5 Hyperexponention

The following series of binary functions was first given by Hilbert [1] p.185: **hyperexponentiation**[**], ordered by **degree**. The binary input in fields 01 and 02 is called **base** and **power**.

| number-constant | description | primitive-arbor-number |
|---|---|---|
| Nmp | multiplication variant x*y | {2,02} {0} |
| (0)Nhxpg | hyperexponentiation degree 0 | {3 {1,0,01} 2} |
| N^ | exponentiation variant $x^y$ | {4,02} {0} 0 {5 |
| (1)Nhxpg | hyperexponentiation degree 1 | {2,02} {0} |
| | | {3 {1,0,01} 2} 4} |
| N^^ | superexponentiation | {6,02} {0} 0 {7 |
| (2)Nhxpg | hyperexponentiation degree 2 | {4,02} {0} 0 {5 |
| | | {2,0 } {0} |
| | | {3 {1,0,01} 2} 4} 6} |
| N^^^ | supersuperexponentiation | {10,02} {0} 0 {11 |
| (3)Nhxpg | hyperexponentiation degree 3 | {6,02} {0} 0 {7 |
| | | {4,0} {0} 0 {5 |
| | | {2,0 } {0} |
| | | {3 {1,0,01} 2} 4} 6} 10} |
| ... | and so on | |

*Table 9. Hyperexponentiation* **primitive-arbor-number** *strings (for binary* **scheme***)*

Looking at the last columns one immediately can see the rule for the generator which works up to a given degree by concatenation of the slightly manipulated (~~2~~ : leave away 2 from 02 ) preceding string in front and at the rear with two **number** strings that follow a simple rule.

It is not difficult to construct a possible number Nhxpg although it may be **very lengthy**, just think of the starting number (in usual decimal notation 828 902 980 983 818 908 901 929 ) for input 0 . You do a loop with the degree as limit and then you concatenate two numbers, one in front, one behind. These pre- and post-numbers follow a simple rule. Concatenation can be built from simple **primitive-arbor-number** strings of table 8 and one needs some rather lengthy constantions like e.g. the above starting number.

$(N_2;N_3)(N_1)$Nhxpg is a *trinary-processive-scheme* with degree $N_1$ , base $N_2$ and power $N_3$ .

In a very natural and simple fashion I get the special case of the Ackermann function that is contained in hyperexponentiation, as given by Hilbert [1] p.185 . When defining these functions with minimisation it is quite complicated. Have you ever seen a textbook where the mimimisation for the Ackermann function has been written down explicitely? Here it is obtained without leaving the concrete calcule NU that is based on primitive recursive functions but obviously allows for much more.

|  | $2 \times N_1$ | $2\hat{}N_1$ | $2\hat{}\hat{}N_1$ | $2\hat{}\hat{}\hat{}N_1$ | $2\hat{}\hat{}\hat{}\hat{}N_1$ |  |
|---|---|---|---|---|---|---|
| degree power | 0 | 1 | 2 | 3 | 4 | ... |
| 0 | 0 | 1 | 1 | 1 | 1 | ... |
| 1 | 2 | 2 | 2 | 2 | 2 | ... |
| 2 | 4 | 4 | 4 | 4 | 4 | ... |
| 3 | 6 | 8 | 16 | 65336 |  | ... |
| 4 | 8 | 16 | 65336 |  |  | ... |
| ... |  |  |  |  |  |  |

*Table 10. Hyperexponentiation lowest values for base* 2

In the preceding section I have said that one has to prove the **Generator** property of a **number** string in every single case.

Theorem B: Existence of a **Generator-number** Nhxpg for hyperexponentiation.
Applied to a **number** string it produces a **primitive-arbor-number** string.

$\forall N_1 [ ((N_1)Mhxpg)Npary = 0 ]$

Proof idea: follows from the very construction

I call a direct proof in a concrete calcule a demonstration; besides that one can prove by deductions too. It is a very intersting question and a big field of future work to find out what rules govern the demonstrations in a concrete calcule.

I have shown an example that via procession one can construct other **scheme** strings within the concrete calcule NU of decimal arbation natural numbers that calculates genuinely recursive functions, e.g.the hyperexponentiation. These **scheme** strings are **complex.**

### 6.6 Hyperincrementation

You may think that the **unary-scheme** as obtained from hyperexponentiation, where base power and degree have the same value $(N_1;N_1)(N_1)$Nhxpg is an extremely fast growing function. But one can do even better.

I introduce the concept with the innocent name of **hyperincrementation**∗∗⁾; it means the series of fastest growth ( incrementatio citissime) given by *unary-scheme* strings.

A rather metaphoric comparison: just as the velocity of light poses an upper limit for all physical motions the consecutive hyperincrementation poses an upper limit for the numbers that can be calculated with a certain expenditure.

The **size** of an *arbor-number* string is defined by the count of fields that appear in it. The size is calculated by a *primitive-scheme* string through a *primitive-arbor-number* string Nsiz. The count of accolades, i.e. matching pairs { } (which includes commas) is equal to **size** if the *number* string ends with character } and the predecessor of **size** if the last digit is an octal number

The idea is: when you look at a *primitive-arbor-number* string you can ask: what is the fastest growth you can produce with a *primitive-arbor-number* string of the same size.

Another measurement in this context is the **depth** given by Ndep that gives the maximum of nested accolades, because in nesting of accolades you get the best explosion rate.

| **number-constant** | *description* | *measurements* | **number** |
|---|---|---|---|
| Ndp (0)Nhicg | duplication 2x degree 0 | size 5 , depth 1 | {0,01} {1,0,01} |
| Nic (1)Nhicg | incrementation $2^x$ x degree 1 | size 9 , depth 2 | {0,01} {1 {2,0} {3,0,2} 01} |
| Nsic (2)Nhicg | superincrementation degree 2 | size 13 , depth 3 | {0,01} {1 {2,0} {3 {4,0} {5,0,4}2} 01} |
| Nssic (3)Nhicg | supersuperincrementation degree 3 | size 17 , depth 4 | {0,01} {1 {2,0} {3 {4,0} {5 {6,0} {7,0,6}4}2} 01} |
| ... | and so on | | |

*Table 11. hyperincrementation primitive-arbor-number strings (for unary-scheme)*

And again I have a relatively simple generator Nhicg for this series, where the members are *unary-scheme* strings that grow eventually faster than any other *unary-scheme* string of the same size. It can be constructed along similar lines as shwown in the preceding section for hyperexponentiation.

Problem: find the series for **nullary hyperincrementation**, i.e. the *primitive-arbor-number* strings that produce the largest output without any input field for a given **size** of the string (it may be a little tricky for small sizes).

### 6.7 Majorant scheme strings

As hyperincrementation functions are the fastest growing functions for a given depth, they allow to determine majorants for all *primitive-scheme* strings.

Theorem C : A certain degree of hyperincrementation provides an eventual majorant for *unary-primitive-scheme* strings

$\forall N_1 [ \exists N_2 [ \exists N_3 [ \forall N_4 [ [ (N_3;N_4)Nemiy = Nn ] \rightarrow [ ((N_4)N_1;(N_4)(N_2)Nhicg )Nemiy = Nn ] ] ] ]$

where Nemiy is the equal-minority-characteristic

Proof idea:

If $N_1$ is not a **primitive-arbor-number** it is trivial, take $N_2 = 0$
Otherwise take $((N_1)Ndepth)N'$ as $N_2$

Problem: find a total majorant for **unary-primitive-scheme** strings (that also takes care of small input values). Using that majorant one can also give a majorant for the count of steps that is needed in the calculation of the values of the **unary-primitive-scheme** string for a value (just multiply by the size of the program **number** ).

### 6.8 Other  *exotic-scheme* strings , especially *paradox-scheme* strings

In the preceding sections there were **processive-scheme** strings like hyperexponentiation and hyperincrementation as examples of **exotic-scheme** strings. As another example of a **processive-scheme** string take the denumeration (with repetitions) of all generated **unary-primitive-scheme** strings with two levels:

A **binary-bis-procession-scheme** string:      $(N_2)((N_1)Ncol)(N_1)Nrow$

and its diagonal                                $(N_1)((N_1)Ncol)(N_1)Nrow$

There are other even more exotic **exotic-scheme** strings in concrete calcule <u>NU</u> of decimal arbation natural numbers, **paradox-scheme** strings that have at least one **number-variable** string in a program position. The simplest example is

the zero value **unary-paradox-scheme**:       $(N_n)N_1$

There are **paradox-scheme** strings that do not contain any **number** or **number-constant** strings. I call them **ex-nihilo-scheme**[*)] strings, they seem to come ex nihilo, from nowhere. Two simple examples are

the diagonal **unary-ex-nihilo-scheme** string:   $(N_1)N_1$
the **trinary-ex-nihilo-scheme** string:          $(N_3)(N_2)N_1$

The above diagonalisations do not lead outside arbative functions as the diagonalisation only relates to a class of arbative functions, the results are proper **scheme** strings.

It is clear that **exotic-scheme** strings do not give primitive recursive functions. What then? Do they correspond to recursive functions? It was shown that in section 6.5. and 6.6 that at least some **processive-scheme** strings like hyperexponentiation and hyperincrementation do. In sections 7.4 and 8.1 I will further discuss this question.

## 7. Metainvestigating arbations

### 7.1 Definition of *phrase*, *sentence* and *formula* strings

In section 6.1 I have defined **scheme** strings that give rise to functions. Now I am going to define **phrase**[*)], **sentence** and **formula**[*)] strings (the latter giving rise to relations).

| | | |
|---|---|---|
| **positive-nullitive-phrase ::** | *pattern* = *nullum-thing* ¦ *nullum-thing* = *pattern* |
| **negative-nullitive-phrase ::** | *pattern* ≠ *nullum-thing* ¦ *nullum-thing* ≠ *pattern* |
| **nullitive-phrase**[**\*)] **::** | *positive-nullitive-phrase* ¦ *negative-nullitive-phrase* |
| **positive-equitive-phrase ::** | *pattern1* = *pattern2* |
| **negative-equitive-phrase ::** | *pattern1* ≠ *pattern2* |
| **equitive-phrase**[**\*)] **::** | *positive-equitive-phrase* ¦ *negative-equitive-phrase* |

The binary metarelation **bound-in** means that the number-variable does appear bound. The binary metarelation **free-in** means that the number-variable appears genuinely, but not bound, and therefore can be bound. **phrase** strings are constructed metarecursively from **equitive-phrase** strings by junctive logic and quantive logic operators:

$$\forall N_1 \, [\, [\, phrase(N_1) \,] \leftrightarrow [\, [\, equitive\text{-}phrase(N_1) \,] \vee$$
$$[\, \exists N_2 \, [\, \exists N_3 \, [\, \exists N_4 \, [\, [\, [\, [\, number\text{-}variable(N_3) \,] \wedge [\, free\text{-}in(N_2; N_3) \,] \,] \wedge$$
$$[\, phrase(N_2) \,] \,] \wedge [\, phrase(N_4) \,] \,] \wedge [\, [\, [\, [\, [\, [\, N_1 = \neg \, [\, N_3 \,] \,] \vee [\, N_1 = [\, N_2 \,] \wedge [\, N_4 \,] \,] \,] \vee$$
$$[\, N_1 = [\, N_2 \,] \vee [\, N_4 \,] \,] \,] \vee [\, N_1 = [\, N_2 \,] \rightarrow [\, N_4 \,] \,] \,] \vee [\, N_1 = [\, N_2 \,] \leftrightarrow [\, N_4 \,] \,] \,] \vee$$
$$[\, N_1 = \forall \, N_3 \, [\, N_2 \,] \,] \,] \vee [\, N_1 = \exists \, N_3 \, [\, N_2 \,] \,] \,] \,] \,] \,] \,] \,]$$

A **phrase** string without a free **number-variable** string is called **nullitive-phrase** or **sentence** string, a **phrase** string with free **number-variable** string is called **formula** string. A **formula** string has an arity that is given by the highest free **number-variable** string.

| | |
|---|---|
| **arithmetic-prog-number-thing ::** | Nnullum ¦ {0} ¦ {0, *input-field*} ¦ |
| | N' ¦ {0,01} 0 ¦ N+ ¦ {0,01}{1,0,02} ¦ N× ¦ {2 {1,0,01} 02} |
| **arithmetic-pattern-array ::** | *arithmetic-pattern* ¦ *arithmetic-pattern-array* ; *arithmetic-pattern* |
| **arithmetic-pattern-argument ::** | ( ) ¦ ( *arithmetic-pattern-array* ) |
| **arithmetic-pattern**[*)] **::** | *number-dingus* ¦ |
| | *arithmetic-pattern-argument*  *arithmetic-prog-number-thing* |

You get **arithmetic-phrase**, **arithmetic-sentence** and **arithmetic-formula** strings if you replace in the above definitions **pattern** by **arithmetic-pattern** and **phrase** by **arithmetic-phrase** accordingly. The **arithmetic-scheme** strings correspond to the **multinomials**; example of a trinary multinomial in traditional notation: $3 + 7 \, x_3 + 2 \, x_1^2 \, x_2 + 53 \, x_1^3 \, x_2^5 \, x_3^6$.

### 7.2 Arithmetic representability of arbative functions

Already the first result of Gödel's famous paper [2] of 1931 was quite surprising: all primitive recursive are arithmetically representable. I can transfer this result immediately to calcule <u>NU</u> :

<u>Metatheorem **B**</u>: All **primitive-scheme** strings of a given arity are arithmetically representable

$$\forall N_1 \, [\, [\, number(N_1) \,] \rightarrow [\, \exists N_2 \, [\, [\, binary\text{-}arithmetic\text{-}formula(N_2) \,] \wedge$$
$$[\, Truth(\, \forall N_1 \, [\, \forall N_2 \, [\, [\, (N_1) N_1 = N_2 \,] \leftrightarrow [\, N_2 \,] \,] \,] \,) \,] \,] \,] \,]$$

$$\forall N_1 \, [\, [\, number(N_1) \,] \rightarrow [\, \exists N_2 \, [\, [\, trinary\text{-}arithmetic\text{-}formula(N_2) \,] \wedge$$
$$[\, Truth(\, \forall N_1 \, [\, \forall N_2 \, [\, \forall N_3 \, [\, [\, (N_1; N_2) N_1 = N_3 \,] \leftrightarrow [\, N_2 \,] \,] \,] \,] \,) \,] \,] \,] \,]$$ and higher arities ...

And as **conventional-scheme** strings can be replaced by **primitive-scheme** it holds

Metatheorem *C:* All **conventional-scheme** strings are arithmetically representable

In the Princeton group [4] [5] of 1936 it was shown that all recursive functions are arithmetically representable too. How about arbative functions, is there a corresponding metatheorem for **exotic-scheme** strings? I can show it immediately for those **processive-scheme** strings that are known to give recursive functions, like e.g. hyperexponentiation.

Metatheorem *D*: Hyperexponentiation is arithmetically representable

$\exists N_1$ *[ [ quaternary-arithmetic-formula($N_1$) ] $\wedge$*
*[ Truth(* $\forall N_1$ *[* $\forall N_2$ *[* $\forall N_3$ *[* $\forall N_4$ *[ [ ($N_2$;$N_3$)($N_1$)Nhxpg = $N_4$ ] $\leftrightarrow$ [ $N_1$ ] ] ] ] ] )] ]*

Proof idea: take the same **arithmetic-formula** string $N_1$ as one has for recursive functions.

So far I have shown that all **conventional-scheme** strings and some **exotic-scheme** strings are arithmetically representable. This brings up the interesting question if all **scheme** strings are arithmetically representable. In section 7.4 I will further discuss this question.

*7.3 Undecidable sentences and the identity problem*

With respect to undecidability theorems and metatheorems there are no changes in concrete calcule NU in comparison to recursive functions: there is no general effective decision procedure. For simplicity I just take the unary case and define **Nully** , **Unnully**\*\*), **Posy**\*\*) and **Unposy**\*\*) strings (for some strange reasons in mathematical logics **Unposy** is called **regular**, although this word is used in other areas of mathematics in some other completely different meanings).:

$\forall N_1$ *[ [ number($N_1$) ] $\rightarrow$*
*[ [ [ [ unary-primitive-Nully($N_1$) ] $\leftrightarrow$ [ Truth(* $\forall N_1$ *[ ($N_1$)$N_1$ = 0 ] )] ] $\wedge$*
*[ [ unary-primitive-Unnully($N_1$) ] $\leftrightarrow$ [ Truth(* $\exists N_1$ *[ ($N_1$)$N_1 \neq$ 0 ] )] ] $\wedge$*
*[ [ unary-primitive-Posy($N_1$) ] $\leftrightarrow$ [ Truth(* $\forall N_1$ *[ ($N_1$)$N_1 \neq$ 0 ] )] ] $\wedge$*
*[ [ unary-primitive-Unposy($N_1$) ] $\leftrightarrow$ [ Truth(* $\exists N_1$ *[ ($N_1$)$N_1$ = 0 ] )] ] ]*

$\forall N_1$ *[* $\forall N_2$ *[ [ [ number($N_1$) ] $\wedge$ [ number($N_2$) ] ] $\rightarrow$*
*[ [ unary-primitive-equality($N_1$;$N_2$) ] $\leftrightarrow$ [ Truth(* $\forall N_1$ *[ ($N_1$)$N_1$ = ($N_1$)$N_2$ ] )] ] ]*

Decidability refers to sentences. At what tier of languages (see section 2.2) does one talk about decidability? A decision for a **sentence** string is a mapping of the **sentence** string to a value true or false. Such a mapping can be performed by a calculation only with respect to numbers that appear in the **sentence** string of a class of **sentence** strings. As two examples:

Primitive decision means that a **primitive-scheme** string is to be evaluated.
Effective decision means that an **effective-scheme**\*) string is to be evaluated.

Theorem D: Primitive undecidability if a **number** string is **unary-primitive-Nully** or **unary-primitive-Unnully** or **unary-primitive-Posy** or **unary-primitive-Unposy** :

$\neg$ *[* $\exists N_1$ *[* $\forall N_2$ *[ [ ($N_2$)$N_1$ = 0 ] $\leftrightarrow$ [* $\forall N_3$ *[ ($N_3$)$N_2$ = 0 ] ] ] ]*

$\neg$ *[* $\exists N_1$ *[* $\forall N_2$ *[ [ ($N_2$)$N_1$ = 0 ] $\leftrightarrow$ [* $\exists N_3$ *[ ($N_3$)$N_2 \neq$ 0 ] ] ] ]*

$\neg \, [ \, \exists N_1 \, [ \, \forall N_2 \, [ \, [ \, (N_2)N_1 = 0 \, ] \leftrightarrow [ \, \exists N_3 \, [ \, (N_3)N_2 = 0 \, ] \, ] \, ] \, ] \, ]$

$\neg \, [ \, \exists N_1 \, [ \, \forall N_2 \, [ \, [ \, (N_2)N_1 = 0 \, ] \leftrightarrow [ \, \forall N_3 \, [ \, (N_3)N_2 \neq 0 \, ] \, ] \, ] \, ] \, ]$

Proof idea: (for the first case **unary-primitive-Nully**, other cases similarily)

Suppose the contrary $\exists N_1 \, [ \, \forall N_2 \, [ \, [ \, (N_2)N_1 = 0 \, ] \leftrightarrow [ \, \forall N_3 \, [ \, (N_3)N_2 = 0 \, ] \, ] \, ]$
Choose $N_2$ as a concatenated string constructed from such an $N_1$ as: $N_2 = N_1 \{2,0\}\{0\}0\{1\{0\}2\}$
This gives the negation of $N_1$
Insert $[ \, (N_1 \{2,0\}\{0\}0\{1\{0\}2\})N_1 = 0 \, ] \leftrightarrow [ \, \forall N_3 \, [ \, (N_3)N_1\{2,0\}\{0\}0\{1\{0\}2\} = 0 \, ] \, ]$
And there is the desired contradiction

Theorem E: Primitively undecidability of **unary-primitive-equality** of **unary-primitive-schemes** strings

$\neg \, [ \, \exists N_1 \, [ \, \forall N_2 \, [ \, \forall N_3 \, [ \, [ \, (N_2;N_3)N_1 = N_n \, ] \leftrightarrow [ \, \forall N_4 \, [ \, (N_4)N_2 = (N_4)N_3 \, ] \, ] \, ] \, ] \, ] \, ]$

Proof idea:
Applying equality characteristic Neqy it can be reduced to the question if a string is **unary-primitive-Nully** as the **unary-scheme** string of the equivalent **equitive-phrase** $((N_4)N_2;(N_4)N_3)Neqy = N_n$ can be replaced by a concatenated **unary-scheme** string $(N_4)N_2 \, N_4 \, N_3 \, N_5$ Neqy where $N_4$ stores the result in a scratch field that is not in $N_3$ and inititalises output field and scratch fields of $N_3$ and $N_5$ puts the stored result of $N_2$ into field 01 and the output of $N_3$ into field 02 and inititalises output field and scratch fields of Neqy .

Let me return to the proof idea for the first case **unary-primitive-Nully** . One could think that the decision procedure of evaluating a **unary-primitive-sentence** string was chosen too simple. There are more possibilities, i.e. **unary-effective-sentence** strings.

**effective-phrase**[*)] strings are constructed metarecursively from **equitive-phrase** strings by junctive[**)] logic and limited quantive[**)] logic operators. Remember $(N_1;N_2)$Nemiy = $N_n$ means $N_1$ less-equal $N_2$

---

$\forall N_1 \, [ \, [ \, \textbf{\textit{effective-phrase}}(N_1) \, ] \leftrightarrow [ \, [ \, \textbf{\textit{equitive-phrase}}(N_1) \, ] \lor$
$[ \, \exists N_2 \, [ \, \exists N_3 \, [ \, \exists N_4 \, [ \, \exists N_5 \, [ \, [ \, [ \, [ \, [ \, \textbf{\textit{number-variable}}(N_2) \, ] \land [ \, \textbf{\textit{term}}(N_3) \, ] \, ] \land$
$[ \, \neg \, [ \, \textbf{\textit{var-free-in}}(N_2;N_3) \, ] \, ] \, ] \land [ \, \textbf{\textit{var-free-in}}(N_2;N_4) \, ] \, ] \land [ \, \textbf{\textit{effective-phrase}}(N_4) \, ] \, ] \land$
$[ \, \textbf{\textit{effective-phrase}}(N_5) \, ] \, ] \land$
$[ \, [ \, [ \, [ \, [ \, N_1 = [ \, \neg \, [ \, N_2 \, ] \, ] \, ] \lor$
$[ \, N_1 = [ \, N_4 \, ] \land [ \, N_5 \, ] \, ] \, ] \lor [ \, N_1 = [ \, N_4 \, ] \lor [ \, N_5 \, ] \, ] \, ] \lor [ \, N_1 = [ \, N_4 \, ] \rightarrow [ \, N_5 \, ] \, ] \, ] \lor$
$[ \, N_1 = \forall \, N_2 \, [ \, [ \, (N_2;N_3)\text{Nemiy} = N_n \, ] \rightarrow [ \, N_4 \, ] \, ] \, ] \lor$
$[ \, N_1 = \exists \, N_2 \, [ \, [ \, (N_2;N_3)\text{Nemiy} = N_n \, ] \rightarrow [ \, N_4 \, ] \, ] \, ] \, ] \, ] \, ] \, ] \, ] \, ] \, ]$

---

The binary metarelation **var-free-in(N;N)** means that the first string appears **properly** free in the second ($N_1$ does not properly appear in $N_{11}$). Every **effective-phrase** string can be calculated effectively for every booking of its **number-variable** strings. A booking is a replacement of **number-variable** strings by **number** strings.

Metatheorem **F**: Effective undecidability if a **number** string is **unary-primitive-Nully** or **unary-primitive-Unnully** or **unary-primitive-Posy** or **unary-primitive-Unposy** , the first case for shortness, where the metafunction **Ninsert(N;N;N)** inserts the third string in the first string whereever the second string appears properly as a **number-variable** string :

$\neg \, [ \, \exists N_1 \, [ \, [ \, \textbf{\textit{unary-effective-scheme}}(N_1) \, ] \land [ \, \forall N_2 \, [ \, [ \, \textbf{\textit{number}}(N_2) \, ] \rightarrow$
$[ \, \textbf{\textit{Truth}}( \, [ \, \textbf{\textit{Ninsert}}(N_1;N_1;N_2) \, ] \leftrightarrow [ \, \forall N_1 \, [ \, (N_1)N_2 = 0 \, ] \, ] \, ] \, ] \, ] \, ]$

Proof idea:
One proceeds in a similar fashion as was used in the last theorem. One reduces the problem to the question if there is an equivalent string for the *unary-effective-scheme* string that is *unary-primitive-Nully*. Some of the tools are already in the toolbox of table 8 in section 4.3 : negation Nneg, conjunction Nand , disjunction Nlor , equality Neqy , inequality Niey , equal-minority Nemiy : implication, biconditional are straightforward and others for limited omnication**[)] and for limited existication**[)] e.g. $\forall N_1 [ [ (N_1;N_2)Nemiy = Nn ] \to$ ... and $\exists N_1 [ [ (N_1;N_2)Nemiy = Nn ] \to$ ... can be added in a fashion similar to Cutland [6] p. 38. Without further discussion:

Metatheorem *E*: Effective undecidability if a *number* string is *primitive-Nully* or *primitive-Unposy*
Metatheorem *F*: Effective undecidability if a *scheme* string is *Nully* or *Unposy*

It will be interesting to further investigate the matter of decidability. For the moment it is enough that the concrete calcule NU has undecidabilities and the **identity problem**. But as decidability depends on the concept of calculability one should always treat that concept first and decidabilty second.

*7.4 Arbative versus recursive functions*
What did I achieve by introducing concrete calcule NU of decimal arbation natural numbers with its arbative functions in comparison to recursive functions? It is visualised in the following figure:

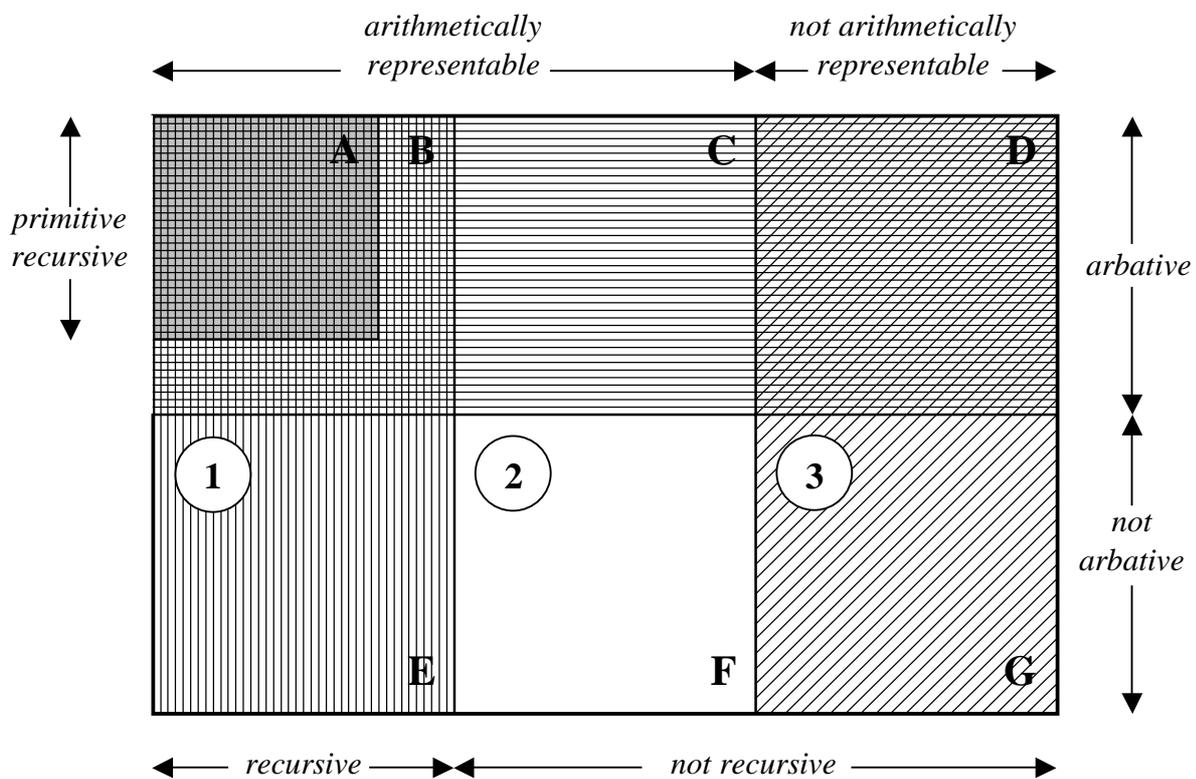

| | | | |
|---|---|---|---|
| **A** | primitive recursive | arithmetically representable | not empty |
| **B** | arbative and recursive (and not primitive recursive) | arithmetically representable | not empty |
| **C** | arbative and not recursive | arithmetically representable | may be empty |
| **D** | arbative and not recursive | not arithmetically representable | may be empty |
| **E** | recursive and not arbative | arithmetically representable | may be empty |
| **F** | extravagant (not recursive and not arbative) | arithmetically representable | may be empty |
| **G** | extravagant | not arithmetically representable | may be empty |

*Figure 4. Set diagram of calculable decimal functions*

In principle there are seven areas but the areas **C**, **D**, **E**, **F** or **G** could be empty. As arbative functions contain all primitive recursive functions area **A** is not empty. As arbative functions e.g. contain hyperexponentiation that is not a primitive recursive function but a recursive function area **B** is not empty either. Areas **F** and **G** will be discussed in section 7.5 ? How about areas **C**, **D** and **E** ?

Church's calculability thesis states that areas **C**, **D**, **F** and **G** are empty. If one of the areas **C** and **D** were nonempty it would mean that Church's calculability thesis is false, as there were calculable functions that are not recursive functions. This conjecture can be formulated as a metasentence about a concrete calcule MU that allows for a precise description of both recursion **and** arbation. In order to talk about both in one calcule one needs two more calculators Mloop(...) and Mhalt(...) that operate for a given number of steps in addition to the calculator Marba(...) , but I cannot discuss it here.

There seems to be a good chance that at least some ***paradox-scheme*** strings can be shown to lead to recursive functions. Perhaps one can adapt the method that Cutland [6] p. 85-99. describes for **universal functions**. Perhaps can simulate the action of the calculator by means of a fix program that I call **interpreter** following the language of applied computing (compare the programming language BASIC in applied computing). Of course the interpreter has to include the check if a ***number*** string is a ***primitive-arbor-number*** string.

In principle exotic functions may appear in areas **C** and **D**. Then the "Diagonal lemma" [7] and Gödel's first incompleteness thesis would need another justification. Otherwise one has to show that exotic functions all belong to area **B**.

If area **E** were nonempty it would mean that there are recursive functions that are not arbative. I doubt it but I cannot prove it. This conjecture can be formulated precisely as a metasentence of the above sketched concrete calcule MU that allows for a precise description of both recursion and arbation.

Suppose it can be shown that areas **C**, **D** and **E** are empty, then recursive and arbative functions are the same. That looks like a nice result, but what would be gained? A lot, as arbative functions are **effectively denumerable** as opposed to recursive functions. This is the cornerstone that following section is built on..

### *7.5 The extravagant Snark-function*

I am aiming for a calculable function that is not arbative. It is obvious that the diagonal method is a sensible choice. Alonso [8a] quotes Kleene that he had realised that one cannot apply the diagonal method for recursive functions: "*When Church proposed this thesis, I sat down to disprove it by diagonalising out the class of the lamda-definable functions. But quickly realising that the diagonalisation cannot be done effectively, I became overnight a supporter of the thesis*".

Although the programs that give rise to recursive functions are effectively denumerable, recursive functions cannot effectively be marked in that series. This is what I have called the ontology problem in section 1.1. It is not a problem that it is an enumeration with repetitions, and it is not a problem that there is the identity problem of section 7.3 (it is not decidable if two programs belong to the same function). It is a problem of marking the good ones. In concrete calcule NU of decimal arbation natural numbers I cannot talk about "all arbative functions" either. However, I can do so effectively in its metacalcule *NU* with the decidable metaproperty ***scheme*** , e.g.

$\forall N_1$ [ [ unary-scheme($N_1$) ] $\rightarrow$ ... and it is a denumeration too: strings are denumerable!

If you want: I can effectively **metatalk** about all arbative functions and therefore I can apply the diagonal method. In order to do that properly I have to provide the tools. I start with

| (idea 8) | **three-tier-multi-level-calculation** as an extreme paradigm. |

In the final passage of section 2.2 I have observed that metacalcules are essentially arithmetic: the strings of a finite alphabet of characters that a metacalcule is talking about can be considered as multal numbers (dual, decimal etc.). This means that one can define all arbative functions for these multal numbers. In addition one can define metarelations and metafunctions with reference to the object-calcule (they start with a capital letter if the involved metaproperty *Truth* is not decidable).

I have defined *unary-scheme* strings of concrete calcule NU. Obviously it is decidable if strings are *unary-scheme* strings. They contain fifteen characters from a **reduced** alphabet of 35 characters of table 5 in section 3.1, as I do not allow the use of comma instead of {} and of *number-constant* strings for the following considerations about a **reduced** concrete calcule NU in the **reduced** metacalcule ***NU***.

| 0 1 2 3 4 5 6 7 { } | 0 1 2 3 4 5 6 7 8 9 | N = ≠ ¬ ∨ ∧ → ↔ ( ; ) ∃ ∀ [ ] |
|---|---|---|
| 0 1 2 3 4 5 6 7 { } | $^1$ | N ( ; ) |

*Table 12. Reduced alphabets for **reduced** concrete calcule NU and for **unary-scheme** strings*

The intrinsic arbative functions of the metacalcule ***NU*** necessitate trigintiquintal ARBACUS and trigintiquintal ARBATOR with programming language **A35**, but that is no problem, actually I just need the primitive recursive functions that are included in arbative functions. All the definitions of metarelations and meta-functions in preceding sections starting from section 3.1 with suffices that small suffices like e.g. *arbor-number* belong to this class, whereas metarelations and metafunctions with suffices that contain capital letters like e.g. ***Posy*** necessitate proofs in the concrete calcule NU of decimal arbation numbers for their evaluation.

For a series of all *unary-scheme* strings I take the series of trigintiquintal numbers in normal ascending order. The trigintiquintal numbers that are not *unary-scheme* strings appear in the series as well and are defined to be equivalent to the nullification *unary-scheme* string {0}, this is of course the vast majority. The procedure is similar to defining *primitive-arbor-number* strings of concrete calule NU, there just are different rules not for *number* strings but the entities of the concrete calcule NU, the trigintiquintal strings. And I remind you that the *number* strings of calcule NU are also strings of the metalanguage.

I can talk about the metaentities of the metalanguage in the metametalanguage. As I do it informally in this paper I use plain English (supralanguage), i.e. the common language instead of a genuine and separate meta-metalanguage.

I have a series of all *unary-scheme* strings, of course with a tremendous amount of trivial ones. When calculating the value of a *unary-scheme* string it is either 0 if it is trivial, or it is calculated for the input *number* by replacing all appearances of the *number-variable* string $N_1$ by this *number* and then the machinery of the concrete calcule NU is started for the actual calculation.

Now I can convert every *unary-scheme* string coded by a trigintiquintal number into a decimal number by a conversion metafunction ***N****trigintiquintal-to-decimal(N)* . The result is a *number* string.

There is the inverse metafunction ***N****decimal-to-trigintiquintal(N)* that converts decimal numbers to trigintiquintal numbers for *number* input, otherwise the result is put to *zero*. The result is a string.

I introduce the metafunction ***N****insert-unary-scheme(N;N)* that puts the second string into the first string at all places instead of all appearances of the string $N_1$ if the first string is a *unary-scheme* string and the second string is a *number* string; otherwise the result is put to *zero* . The result is a *term* string.

I introduce the metafunction ***N****successor(N)* that calculates its successor of the string, a trigintiquintal number. The result is a string, the next trigintiquintal number (succession is in metacalcule ***NU*** ).

It is immediately clear that all auxiliary metafunctions that are used in constructing it are total and primitive recursive with respect to the metacalcule *NU* .

I finally introduce the metafunction *NEvaluate(N)* that calculates the value of a a **term** string, if the input is a **term** string, otherwise the result is put to **zero** . Therefore the result is always a **number** string. The calculation is done according to the rules of the concrete calcule NU of decimal arbation natural numbers. This metafunction is not a primitive recursive with respect to the metacalcule *NU* but is precisely defined like follows (watch out for the different fonts of *=* and = ):

$\forall N_1$ *[*$\forall N_2$ *[*
*[ [ [ term(*$N_1$*) ] ∧ [ number(*$N_2$*) ] ] → [ [ NEvaluate(*$N_1$*)* = $N_2$ *] ↔ [ Truth(* $N_1$ = $N_2$ *) ] ] ] ∧*
*[ [ ¬ [ [ term(*$N_1$*) ] ∧ [ number(*$N_2$*) ] ] ] → [ NEvaluate(*$N_1$*)* = *nullum ] ] ]*

Nobody can keep me away from evaluating the series of all **unary-scheme** strings for their own value, that has been converted into a decimal **number** string and take its successor. You see: classical diagonalisation producing a unary metascheme *NBoojum(*$N_1$*)*

$\forall N_1$ *[NBoojum(*$N_1$*) = Nsuccessor(*
*Ndecimal-to-trigintiquintal(NEvaluate(Ninsert-unary-scheme(*$N_1$*;Ntrigintiquintal-to-decimal(*$N_1$*) ) ) ) ]*

From this unary metascheme I construct an extravagant*[)] function that I call the **Snark**-function*[)]. The Snark-function is defined for all decimal numbers, the result is always a decimal number. It is obtained by translating the argument decimal number into a trigintiquintal number. Then the metafunction *NBoojum(N)*[*)] is calculated for this trigintiquintal number, the result is translated back into a decimal number (decimal is necessary for comparison in figure 4). The Snark-function is not in the metacalcule NU , but is only describable in the third tier, the supralanguage. But you can also use the expression "calculable" only in the third tier, the supralanguage.

The metafunction *NBoojum(N)* is not contained in the **unary-scheme** string series due to classical diagonalisation: therefore it is not an arbative function and the derived Snark-function is not an arbative function either. The important point is that it is calculable! It is sort of a new "Ackermann" function. How is the Snark-function to be placed. in figure 4 ? There are three possibilities:

**1** recursive (and thus arithmetically representable)      in area **E**
**2** arithmetically representable, but not recursive      in area **F**
**3** not arithmetically representable and not recursive      in area **G**

With possibilty **1** Church's calculability thesis could survive. With possibilty **2** Gödel's calculability thesis could survive. But for me there seems to be little chance that the Snark-function is arithmetically representable. As long as nobody shows that the Snark-function is arithmetically representable possibilty **3** cannot be ruled out with the consequence that the "Diagonal lemma" and Gödel's first incompleteness thesis are in **limbo**.

*Remark: If somebody could show that the Snark-function is arithmetically representable, there still remain doubts as to Gödel's first incompleteness thesis. There may be many concrete arithmetical calcules, metacalcules thereof and even calcules of higher tier that embrace recursive functions and have more functions. Maybe these calcules cannot be combined into a single one. One could even think of an infinite ladder of metacalcules, where you take one function out of every ladder-step. Or other wild things in metacalcule NU like* 0    (1)1    (2)(2)2    (3)(3)(3)3    *and so on. Who knows about the fantasy of mathematicians and metamathematicians. There is not necessarily a "mother of all calculators". And somebody could come up with another proposal for another weird function that has yet to be shown to be arithmetically representable.*

## 8. Conclusion

### 8.1 Challenging the defenders of Church's calculability thesis

If someone has formulated a thesis (or a conjecture) that all elements have a certain property, and someone puts forward a special precisely defined element it is the responsibilty of the defender of that thesis (or that conjecture) to show that this special element has this certain property. As long as it is not shown that thesis (or that conjecture) is in limbo. Such is the logic of a thesis (or a conjecture).

Church's calculability thesis states that all calculable functions are recursive functions. I have put forward arbative functions. Arbative functions are calculable. As they are given by the **scheme** strings of concrete calcule <u>NU</u> of decimal arbation natural numbers they can be effectively denumerated in its metacalcule <u>*NU*</u>. Therefore it is up to the defenders of Church's calculability thesis to show that all arbative functions are recursive.

<u>Challenge 1.1</u>: the defenders of Church's calculability thesis have to show that areas **C** and **D** of figure 4 are empty, meaning that there are no arbative functions that are not recursive. All exotic arbative functions have to be shown to be recursive, meaning that they all belong to area **B**.

The challenge is formulated in a precise manner. As long as it is not answered in a correspondingly precise manner Church's calculability thesis is in limbo. The method of Cutland [6] p. 85-99 for universal functions may be a good starting point to meet this challenge.

<u>Challenge 1.2</u>: the defenders of Church's calculability thesis have to show that the Snark-function is recursive, meaning that possibilty **1** of figure 4 applies. No problems would then arise for Gödel's first incompleteness thesis.

### 8.2 Challenging the defenders of Gödel's calculability thesis

Gödels's calculability thesis is weaker than Church's. So there is less to show:

<u>Challenge 2.1</u>: the defenders of Gödels's calculability thesis have to show that area **D** of figure 4 is empty, i.e. that all exotic functions are arithmetically representable

<u>Challenge 2.2</u>: the defenders of Gödels's calculability thesis have to show that possibilities **1** or **2** of figure 4 apply, i.e. that the Snark-function is arithmetically representable.

No problems would then arise for Gödel's first incompleteness thesis.

I do not advise to put to much work right now into meeting challenge 2.2. The results of section 7.5 has given me the courage for a forthcoming publication: I will present an abstract calcule of natural numbers, that seems to have a **categorical** set of axioms (see section 2.4), which would contradict some theorems of Skolem [3] and Gödel's first incompleteness thesis; the latter implies that there is no axiomatic system that completely describes natural numbers. Skolem stands for the large majority of mathematicians that adhere to a certain Platonism without which a lot of mathematics would not be possible but that has to be questioned when it comes to the matter of "effective" calculations. A critical analysis of Skolem's theorems is necessary.

### 8.3 Counterchallenge welcome

Finally I put forward the **conjecture** that area **E** of figure 4 is empty meaning that there are no recursive functions that cannot be expressed as arbative functions. If on looks at unary functions only: is there a **unary-scheme** string for every unary non-primitive recursive function (i.e. one with at least one minimisation)? I do think so but I cannot prove it. But in the sense of the introductory remark of the preceding section I accept as

Counterchallenge: if I am presented with a non-primitive recursive function, I have to show that I can write down the appropriate **scheme** string of the concrete calcule NU of decimal arbation natural numbers that gives rise to this function. As it all comes down for the challenger to show that minimisation functions are "regular" I am quite confident, because that proof has to be presented to the defender. And I think that one can obtain the recipe for the **scheme** from an analysis of that proof.

### *8.4 Résumé*
Why did the Princeton group invent minimisation for the inclusion of Ackermann functions, why didn't they go a way like the one that has been sketched in this publication? I think it is so much easier today: now there are real computers, program languages, program codes and program generators. Since some years there is the idea of hypercomputation [8b], however, it tries to tackle the problem straight on and has not yet been overall successfull. I have chosen a bypath.

Suppose that the challenges 2.1 and 2.2 are met and the Snark-function turns out to be arithmetically representable, then Gödel's first incompleteness thesis would survive, but there still would be the challenges 1.1 and 1.2. If these were also met Church's calculability thesis would survive. But in any case, besides a new way to look at computers and calculators the following would remain of my reasoning:

*For a logical analysis of a sentence one should always ask to what tier of languages it belongs and try to write it down in the appropriate language. And one never should call a thesis or a conjecture a theorem.*

**Table of words with special meaning and newly coined expressions** (pages of first appearance)